\documentclass[a4paper,11pt]{article}
\pdfoutput=1
\usepackage[no-natbib-sort]{jheppub}

\usepackage[utf8]{inputenc}
\usepackage{etoolbox}
\usepackage{hhline}
\usepackage{array}
\usepackage{amssymb,amsmath,amsthm,amsfonts,graphicx}
\usepackage{latexsym}
\usepackage{bm}
\usepackage[]{hyperref}
\usepackage{cleveref}
\usepackage{nccmath}	
\usepackage{color}
\usepackage{amssymb,amsbsy,enumitem}

\makeatletter
    \patchcmd{\maketitle}{\@fpheader}{}{}{}
    \makeatother

\def\be{\begin{equation}}
\def\ee{\end{equation}}
\def\ba{\begin{eqnarray}}
\def\ea{\end{eqnarray}}

\title{The origin of the Reissner-Nordstr\"om de Sitter instability}

\author[a]{Oscar~J.~C.~Dias,}
\emailAdd{ojcd1r13@soton.ac.uk}
\affiliation[a]{STAG research centre and Mathematical Sciences, Highfield Campus, University of Southampton, Southampton SO17 1BJ, UK}

\author[b,c]{and Jorge~E.~Santos}
\emailAdd{jss55@cam.ac.uk}
\affiliation[b]{Department of Applied Mathematics and Theoretical Physics, University of Cambridge, Wilberforce Road, Cambridge CB3 0WA, UK} 
\affiliation[c]{Institute for Advanced Study, Princeton, NJ 08540, USA}

\abstract{We give strong numerical evidence for the existence of an instability afflicting six-dimensional Reissner-Nordstr\"om de Sitter (RNdS) black holes. This instability is akin of the Konoplya-Zhidenko instability present in RNdS black holes in seven spacetime dimensions and above. Moreover, we perform a detailed analysis of the near-horizon limit of extremal RNdS black holes, and find that unstable gravitational modes effectively behave as a massive scalar field whose mass violates the AdS$_2$ Breitenl\"ohner-Freedman bound (if and only if $d\geq 6$), thus providing a physical argument for the existence of the instability. Finally, we show that the frequency spectrum of perturbations of RNdS has a remarkable intricate structure with several bifurcations/mergers that appears unique to RNdS black holes.}

\begin{document}

\maketitle

\section{Introduction}

Kodama and Ishibashi proved that Reissner-Nordstr\"om de Sitter (RNdS) black holes in $d=4$ and $d=5$ spacetime dimensions are linearly-mode stable \cite{Kodama:2003kk}. Therefore, the finding by Konoplya and Zhidenko  \cite{Konoplya:2008au,Konoplya:2013sba}, that RNdS black holes can be unstable to gravitational perturbations if they live in $d\geq 7$ spacetime dimensions ($n=d-2\geq 5$) came with some surprise. The existence of this instability was further confirmed in \cite{Cardoso:2010rz} where it was also noted that there is a simple (necessary but not sufficient) criterion $-$ originally due to \cite{Buell:1995} $-$ that predicts the existence of the instability. Essentially,  \cite{Buell:1995} proposes that a system should be unstable whenever an integral (between the event and cosmological horizons) of the Schr\"odinger potential of the perturbation is negative. The instability of   \cite{Konoplya:2008au} is in these conditions \cite{Cardoso:2010rz}.  More recently, this RNdS instability was also studied in the framework of the large $d$ limit of general relativity \cite{Tanabe:2015isb}.

There are however some fundamental questions that are left unanswered by \cite{Konoplya:2008au,Cardoso:2010rz,Konoplya:2013sba,Tanabe:2015isb} and that we address in this manuscript. Firstly, we would like to understand the physical origin of this instability, and in particular why it only appears in higher-dimensions. This is particularly important, since for $d=4$ and $d=5$ Kodama and Ishibashi proved linear-mode stability \cite{Kodama:2003kk}. It is also important in the context of recent studies on strong cosmic censorship violation in RNdS black holes \cite{penrose,Cardoso:2017soq,Costa:2017tjc,Dias:2018ynt,Dias:2018etb,Dafermos:2018tha,Hod:2018dpx,Dias:2018ufh,Dias:2019ery,Zhang:2019nye,Gim:2019rkl,Liu:2019lon}. Secondly,  \cite{Konoplya:2008au,Cardoso:2010rz} find that the system is unstable for $d\geq 7$ but their analysis leaves the stability properties of the $d=6$ case undetermined. Again, stability is proven only for $d=4,5$ so could it be that there is an instability also for $d=6$?

In this paper we address these two questions. In section \ref{sec:background} we start by reviewing the RNdS black holes and its relevant sector of perturbations that can be unstable. Then, in section \ref{sec:NH} we point out that there is a criterion for instability that predicts and, at the same time justifies, an instability in RNdS black holes. This instability criterion was conjectured by Durkee and Reall \cite{Durkee:2010ea} and later proved by Hollands and Wald \cite{Hollands:2014lra}. Strictly speaking it was proven only for non-positive cosmological constant backgrounds, $\Lambda\leq 0$, but, as argued in \cite{Hollands:2014lra}, it should also hold for de Sitter backgrounds ($\Lambda>0$).  Applied to our system, this Durkee-Reall instability criterion essentially states that that one should take the near-horizon limit of the gravitational perturbation master equation about the extremal RNdS black hole. After the limit is taken, the gravitational master equation effectively reduces to a Klein-Gordon equation for a massive scalar field in an AdS$_2$ background. If this near-horizon effective mass is smaller than the  AdS$_2$ Breitenl\"ohner-Freedman (BF) mass bound then the full RNdS extremal geometry should be unstable. By continuity this instability should extend away from extremality. As shown in section \ref{sec:NH}, this criterion predicts the existence of an instability for RNdS black holes in $d\geq 6$. In section \ref{sec:results}, we numerically solve  the perturbation master equation and we  confirm that the instability is indeed present for  RNdS black holes in $d\geq 6$ dimensions.  Therefore, the two main outcomes of our study  are: 1) we provide a physical origin of the Konoplya-Zhidenko instability (violation of the AdS$_2$ BF bound), and 2) we establish the presence of the instability in $d=6$. 

Adding to these two main results, we will also take the opportunity to explore in detail the properties of the instability. 
For example, we will establish that the instability is present in a broader region of the parameter space than originally reported in \cite{Konoplya:2008au,Cardoso:2010rz,Konoplya:2013sba}. Indeed, in one of our studies we will look for the instability directly in the extremal configuration and will find that {\it all} extremal RNdS black holes are unstable for $d\geq 6$\footnote{For $d=6$, the evidence is substantially weaker, as the instability growth rates are very small.}. That is,  \cite{Konoplya:2008au,Cardoso:2010rz,Konoplya:2013sba}  identified the presence of the instability only for large values of $r_+/r_c$ (where $r_+$ and $r_c$ are the event and cosmological horizons of RNdS) but we find that the instability is actually present in the whole range $r_+/r_c\in ]0,1[$ for extremal RNdS. Our findings are not in conflict with the Durkee-Reall instability criterion, since the latter is known to be a sufficient, but not necessary condition for the existence of an instability. Additionally, we will also analyse the properties of the instability away from extremality and we will also look directly for the onset of the instability. Finally, we will also study in some detail the quasinormal mode structure of the perturbations as we span the two-dimensional parameter space of RNdS black holes. We will find an intricate network of quasinormal mode branches with interesting bifurcations/mergers that, to the best of our knowledge, seem to be unusual in black hole perturbations (at least in $\Lambda\leq 0$ backgrounds).

\section{Reissner-Nordstr\"om de Sitter black hole and its perturbations}
\label{sec:background}

We work with the Einstein-Maxwell theory,  in $d=n+2$ spacetime dimensions ($n\geq 2$), with a positive cosmological constant $\Lambda$ described by the action
\be
S=\frac{1}{16\pi G}\int \mathrm{d}^{n+2} x \sqrt{-g}\left(R -2\Lambda -F^2 \right)\,, \quad \hbox{with} \quad \Lambda \equiv \frac{n(n+1)}{2L^2}
\ee
and where $R$ is the Ricci scalar of the metric $g$, $L$ is the de Sitter length scale, and $F=\mathrm{d}A$ is the Maxwell field strength associated with the Maxwell potential $A$.

A known solution of this theory is the Reissner-Nordstr\"om de Sitter (RNdS) black hole.
In static coordinates, the gravitational and electric fields of this solution with mass $M$ and charge  $q$ parameters  are
\be
\mathrm{d}s^2=-f\,\mathrm{d}t^2+\frac{\mathrm{d}r^2}{f}+r^2\mathrm{d}\Omega_n^2\,,\qquad A=-\frac{q}{r}\mathrm{d}t\,,
 \label{metricRN}
\ee
with $\mathrm{d}\Omega_n^2$ being the line element of a unit radius $S^n$ and 
\be
f(r)=1-\frac{r^2}{L^2} -\frac{2M}{r^{n-1}}+\frac{Q^2}{r^{2(n-1)}}\,, \qquad Q=\frac{\sqrt{2}\, q}{\sqrt{n (n-1)}}\,.
\label{metricRNaux}
\ee

For an appropriate range of parameters, specified below, $f$ has three real positive roots $r_-\le r_+\le r_c$ corresponding to the Cauchy horizon $\mathcal{CH}$, event horizon  $\mathcal{H}^+$  and cosmological horizon $\mathcal{H}_C$ respectively. We can express $M$ and $L$ in terms of $r_+$, $r_c$ and $q$. The temperature of the event and cosmological horizons are, respectively, given by $T_+=\frac{f'(r_+)}{4\pi}$ and $T_c=-\frac{f'(r_c)}{4\pi}$.

When $T_+=0$ we have an extremal RNdS black hole. This happens for $q=q_{\rm ext}$ with
\be
\frac{q_{\rm ext}}{ r_c^{n-1}}= y_+^{n-1} \, \sqrt{\frac{n (n-1)}{2}} \sqrt{\frac{2 y_+^{n+1}-(n+1)y_+^2+n-1}{(n+1) y_+^{2 n}-2 n y_+^{n+1}+n-1}}\,,\qquad \hbox{and}\quad y_+\equiv \frac{r_+}{r_c}\,.
\label{Qext}
\ee
The Einstein-Maxwell equations of motion are invariant under the scaling $g\to \lambda^2 g$, $A\to\lambda A$ and $L\to\lambda L$, with $\lambda\in\mathbb{R}$, which we can use to construct dimensionless quantities in units of $r_c$. Therefore, we choose to parametrize the RNdS solution using the dimensionless parameters $q/q_{\rm ext}$ and $y_+$. 

We are interested on gravitoelectromagnetic perturbations of RNdS. These were studied in detail by Kodama-Ishibashi (KI) in \cite{Kodama:2003kk}. Perturbations of (\ref{metricRN}) can be analysed according to how they transform under diffeomorphisms of the $S^n$ sphere. There are a total of three families of perturbations that decouple from each other, namely the scalar, vector and tensor perturbations. These perturbations are built from scalar, vector and tensor harmonics on $S^n$, respectively. We are primarily interested in scalar perturbations, which are built from spherical harmonics $S_\ell(\vec{x})$, where $\vec{x}$ collectively parametrise coordinates on the $n-$ sphere. These harmonics are such that
\begin{equation}
\Box_{S^n}S_\ell(\vec{x})=\lambda_{S}S_\ell(\vec{x})
\end{equation}
with $\lambda_S=\ell(\ell+n-1)$ and $\ell\geq0$ being an integer. Modes with $\ell=0,1$ were shown to be pure gauge in \cite{Kodama:2003kk}. In particular, modes with $\ell=0$ describe changes in the mass of the background RNdS black hole while $\ell=1$ modes represent translations. Onwards, we shall take $\ell\geq2$.

In the Schwarzschild limit, $q=0$, $\Phi_{\ell}^{{\rm S}\;-}(t,r)$ and $\Phi_{\ell}^{{\rm S}\;+}(t,r)$ describe, respectively, purely gravitational and purely electromagnetic perturbations \cite{Kodama:2003kk}. However, when $q\neq 0$ the gravito-electromagnetic perturbations are coupled. As described in \cite{Kodama:2003kk}, one can introduce a separation {\it anstaz} of the form 
\begin{equation}
\Phi_{\ell}^{{\rm S}\;-}(t,r)=e^{-i\omega t}S_\ell(\vec{x})\Phi^{-}_{\omega \ell}(r)
\end{equation}
which introduces  the frequency $\omega$. One can then manipulate the Einstein and Maxwell equation to find a decoupled ordinary differential equation for the radial master field $\Phi^{-}_{\omega \ell}(r)$ (see Eq.~(5.59) of \cite{Kodama:2003kk}), namely:
\begin{equation}
\label{KImaster} 
f\left(f \Phi_{\omega \ell}^{-}(r)^{\prime} \right)^\prime +\left( \omega^2 -V_- \right)\Phi^{-}_{\omega \ell}(r)=0 \,,
\end{equation}  
where the potential $V_-(r;n,r_+,r_c,q,\ell)$ can be found in equations (5.61)-(5.63) of \cite{Kodama:2003kk}.

\section{Near-horizon  criterion for instability}
\label{sec:NH}

As reviewed below, the near-horizon limit of an extremal RNdS black hole is described by the direct product spacetime AdS$_2\times S^n$. One of the main observations of the current manuscript is that the existence of some instabilities of the full RNdS black hole can be inferred from studying the behaviour of the  perturbation equation in the near horizon limit. More concretely, in the near-horizon limit the Kodama-Ishibashi master equation \eqref{KImaster}  reduces to an equation that has the form of  a Klein-Gordon equation for a massive scalar field in AdS$_2$. We will confirm that, according to the Durkee-Reall criterion \cite{Durkee:2010ea,Hollands:2014lra}, when this near-horizon effective mass  violates the AdS$_2$ Breitenl\"ohner-Freedman mass bound then the full RNdS geometry is unstable.\footnote{Strictly speaking the Durkee-Reall conjecture was originally formulated and proved only for $\Lambda\geq 0$ backgrounds \cite{Durkee:2010ea,Hollands:2014lra} but, as argued in \cite{Hollands:2014lra} it should also hold for $\Lambda>0$.} This happens for $d\geq 6$ ($n\geq 4$). We will find that the AdS$_2$ BF bound violation gives a sufficient (but not necessary) criterion for the presence of an instability and also justifies its origin.

To get the near-horizon geometry of the extremal RNdS  black hole, one first takes \eqref{metricRN} with $q=q_{\rm ext}$ and  zooms in around the event horizon region by making the coordinate transformations:
\begin{eqnarray} \label{NHtransf}
&& r = r_{+} + \varepsilon \rho, \qquad
t = L_2^{2} \frac{\tau}{\varepsilon}\,;
\nonumber \\
&& \hbox{with} \quad  L_2^2=\frac{r_+^2 \left[2 n y_+^{n+1}-(n+1) y_+^{2 n}-n+1\right]}{(n-1) \left[-4 n y_+^{n+1}+(n+1) \left(y_+^{2 n}+n y_+^2\right)-(n-1)^2\right]}.
\end{eqnarray}
The near-horizon solution is then obtained by taking $\varepsilon \to 0$  yielding (after a $U(1)$ gauge transformation)
\begin{eqnarray} \label{NHgeometry}
&& ds_{NH}^{2} = L_2^{2}\left(-\rho^{2} d\tau^{2} + \frac{d\rho^{2}}{\rho^{2}}\right) + r_{+}^2 d \Omega_{n}^{2},   \nonumber \\
&& A_{\mu}^{NH} dx^{\mu} = \alpha \,\rho \, d\tau\,, \qquad  \alpha =\frac{n-1}{\sqrt{2}} L_2 \sqrt{1 + (n-1)\frac{L_2^{2}}{r_{+}^{2}}}\,.
\end{eqnarray}
This geometry is the direct product of $AdS_2\times$S$^n$ and has a Maxwell potential that is linear in the radial direction. This limiting solution is still a solution of the $(n+2)$-dimensional Einstein-Maxwell-dS theory. On the other hand, the AdS$_2$ metric solves the 2-dimensional Einstein-AdS equations with $AdS_{2}$ radius $L_2$, $R_{\mu\nu}=-L_2^{-2} g_{\mu\nu}$.

One can now take the near-horizon limit \eqref{NHtransf}, together with $\omega=\varepsilon  L_2^{-2} \tilde{\omega}$  directly on the gravitational master equation \eqref{KImaster}. After taking $\varepsilon \to 0$ this yields
\begin{equation} 
 \Big(  \Box_2 - \mu^2L_2^2 \Big) \Phi^{-}_{\omega \ell}(\rho)=0  \,,
\label{master:scalar} 
\end{equation}  
where $\Box_2$ is the d'Alembertian in  AdS$_2$  and $\mu^2=\mu^2(n,y_+,\ell)$ is the near-horizon effective mass of the system. Its particular expression is long and not enlightening so we do not reproduce it here. It depends on the harmonic number $\ell$, on the dimension $d=n+2$ and on $y_+$. The keypoint is that this near-horizon analysis is expected to provide a criterion for instability \cite{Durkee:2010ea,Hollands:2014lra}: whenever this mass is smaller than the AdS$_2$  Breitenl\"ohner-Freedman (BF), $\mu^2_{\rm BF}L_2^2=-1/4$,  one should have an instability of the full RNdS black hole. We further expect this instability to be the one found in \cite{Konoplya:2008au,Cardoso:2010rz,Tanabe:2015isb} (for $d\geq 7$). If so, the violation of the AdS$_2$ BF bound effectively explains the origin of the instability found in \cite{Konoplya:2008au,Cardoso:2010rz,Tanabe:2015isb}. 

 \begin{figure}[b]
	\centering 	\includegraphics[width=0.6\textwidth]{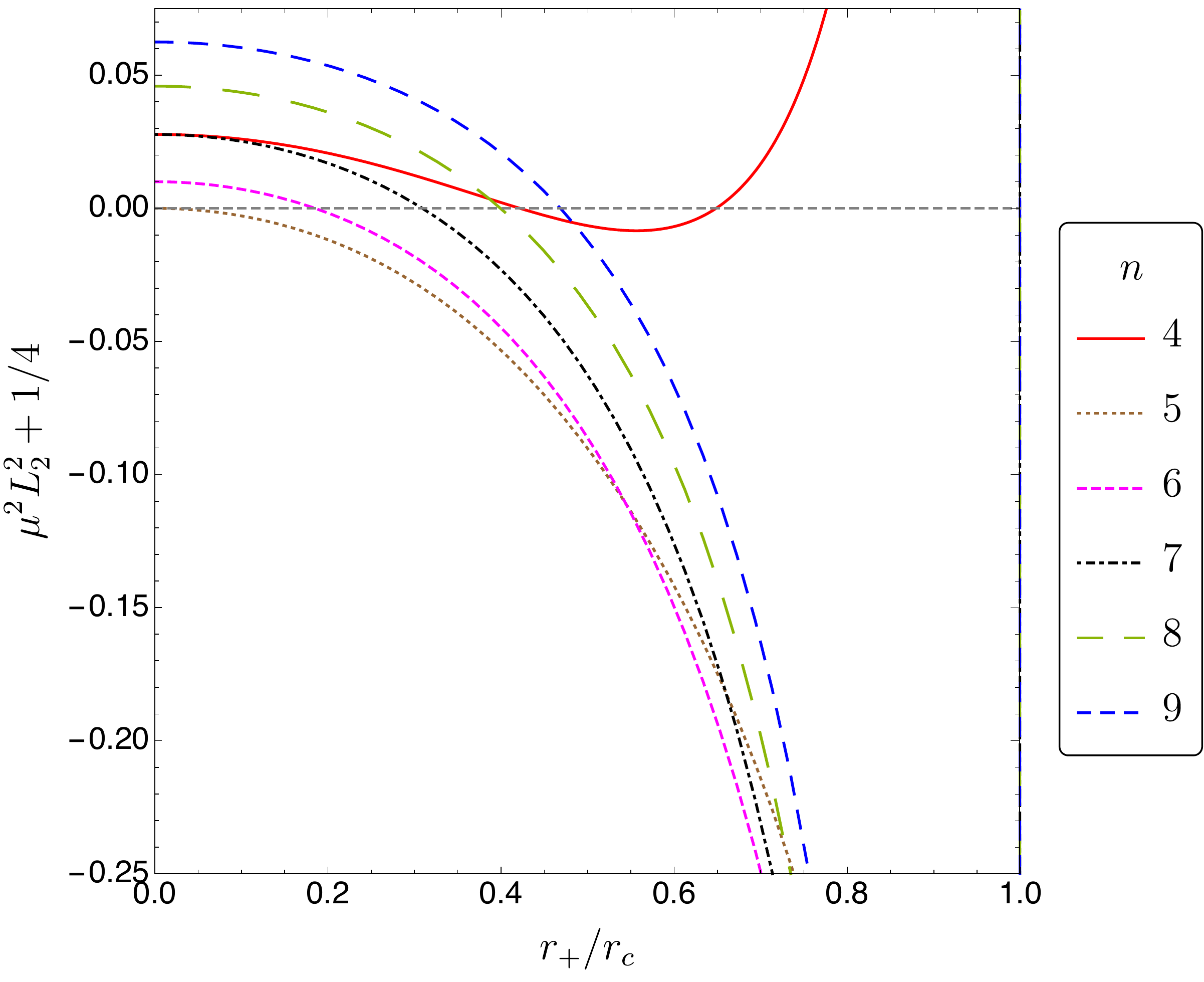} 
		\caption{Difference between the effective near-horizon AdS$_2$ mass $\mu^2 L^2_{2}$ and the AdS$_2$ BF bound $\mu^2_{BF} L^2_{AdS_2}=-1/4$ as a function of $r_+/r_c$. When this quantity is negative one should expect an instability \cite{Durkee:2010ea,Hollands:2014lra}.} 
		\label{fig:NH}
 \end{figure} 
 
In Fig. \ref{fig:NH} we set $\ell=2$ and plot $\left(\mu^2-\mu^2_{\rm BF}\right) L^2_{2}$ as a function of $y_+=r_+/r_c$ for  $n=4,5,6,7,8,9$ (see plot legends). We conclude that for $n\geq 4$ there are always values of $y_+$ where $\left(\mu^2-\mu^2_{\rm BF}\right) L^2_{2}<0$ and thus for which the AdS$_2$ BF bound is violated and an instability should be present. Note that this includes the $d=6$ ($n=4$) case, which was also analysed in \cite{Konoplya:2008au,Cardoso:2010rz}, but for which no instability was found. For $n=2,3$, i.e. $d=4,5$, one always has $\mu^2> \mu^2_{\rm BF}$ and we do not display these curves in  Fig. \ref{fig:NH}. This result is consistent with the fact that $d=4,5$ RNdS are stable to all linear-mode gravitational perturbations \cite{Kodama:2003kk}.

 In the next section we will confirm that the AdS$_2$ BF bound violation observed in  Fig. \ref{fig:NH}  (for $n\geq 4$) indeed gives a sufficient criterion for the presence of an instability and also justifies its origin. But  we will also show that this condition is not a necessary condition.

So far we discussed only harmonics with $\ell=2$. But we can also ask what happens if we consider modes with $\ell\geq 3$.
 Typically, we find that as the integer $\ell$ increases it becomes harder to get negative  $\left(\mu^2-\mu^2_{\rm BF}\right) L^2_{2}$. For example, in $n=4$ or $n=5$ dimensions, for $\ell\geq 3$ the AdS$_2$ BF bound is no longer violated for any $y_+$. As another example, for $n=6,7$ the harmonics $\ell=2,3$  generate a violation of the AdS$_2$ BF bound but this is no longer the case for $\ell\geq 4$. As a final example,  for $n=8,9$ the harmonics $\ell=2,3,4$  generate a violation of the AdS$_2$ BF bound but this is no longer the case for $\ell\geq 5$. 
 
\section{Numerical results for the instability} 
\label{sec:results} 

\subsection{Setup of the problem} 
\label{sec:results1} 

Our aim is to solve the KI linear ODE \eqref{KImaster}, subject to relevant boundary conditions, to find the properties of linear mode perturbations in RNdS. More concretely, we want to fix the spacetime dimension $n$  and the RNdS black hole $-$ described by the dimensionless quantities $\{y_+,q/q_{\mathrm{ext}}\}$  $-$ as well as the harmonic quantum number $\ell$ and look for modes that are unstable.

To achieve our aims we have followed a strategy that is anchored on three main studies:

\begin{enumerate}[label=\Roman{*}.]

\item Generically, we have a non-extremal RNdS black hole with $q/q_{\mathrm{ext}}<1$. We can  solve \eqref{KImaster} as a quadratic eigenvalue problem for $\omega$ to find the eigenvalue/eigenfunction pair $\left(\omega r_c,\Phi^{-}_{\omega \ell}\right)$ for a given set of $\{n,y_+,q/q_{\mathrm{ext}},\ell\}$. For a fixed $\{n,\ell\}$ we can then repeat the analysis and scan the full 2-dimensional parameter space of RNdS black holes.  
\item As argued in section \ref{sec:NH} the instability has a near-horizon/extremal origin. So, when present, it should make its appearance at extremality and then extend away from extremality until it eventually shuts down. Therefore, we first consider the RNdS with $q=q_{\mathrm{ext}}$ and solve \eqref{KImaster} as a quadratic eigenvalue problem for the frequency to find the eigenpair $\left(\omega r_c,\Phi^{-}_{\omega \ell}\right)$, for a given $\{n,y_+,\ell\}$ directly at extremality.
\item Alternatively, instead of looking for the instability timescale, we can search directly for the onset of the instability whereby the frequency vanishes, $\omega=0$. To find this instability threshold, we solve \eqref{KImaster} as a nonlinear eigenvalue problem for the black hole charge to find the onset charge $q=q^{\mathrm{onset}}$ above which RNdS is unstable.
\end{enumerate}

The above three strategies are complementary. In particular, the 2-dimensional surface in the plot $\{y_+,q/q_{\mathrm{ext}},{\rm Im}(\omega r_c) \}$ generated using the first study must: i) have the extremal curve with $q/q_{\mathrm{ext}}=1$ obtained in the second study as a boundary line, and ii) intersect the onset curve $\omega=0$ of the third study when ${\rm Im}(\omega r_c)=0$. So the three complementary studies also provide  non-trivial independent checks of the numerical results we obtain. 

To proceed, one needs to use a numerical scheme to solve our boundary value problems. For that it is good to first introduce a new compact radial coordinate, namely
\be
y = \sqrt{1-\sqrt{\frac{r_c-r}{r_c-r_+}}}\,,
\ee
such that $y=0$ describes the black hole horizon, $r=r_+$,  and $y=1$ marks the location of the cosmological horizon,  $r=r_c$. 

Next, one needs to discuss the issue of the boundary conditions. These boundary conditions are different for the three eigenvalue problems I$-$III listed above and we discuss them separately. 

Consider first the eigenvalue problem I. We have a non-extremal black hole and deformations propagate between the event and cosmological horizons where we impose as boundary conditions that  the perturbations are regular in Eddington-Finkelstein coordinates. In particular, we take the modes to be ingoing at the event horizon and outgoing at the cosmological horizon. To find these boundary conditions we first do a Frobenius analysis about $y=0$ to find that  
\be
\Phi^{-}_{\omega \ell}(y)\simeq y^{\pm\,\frac{i\,\tilde{\omega}}{2\pi \tilde{T}_+}}\big[1+\mathcal{O}(y)\big]\,,
\label{NonExt:BCeventH}
\ee
where $T_+$ is the event horizon temperature (its surface gravity divided by $2\pi$) as defined below \eqref{metricRNaux} and the tilde (here and in other expressions) is used to state that the quantities are measured in units of $r_c$, {\it e.g.},
\be
\tilde{\omega}\equiv\omega\,r_c \,, \qquad \tilde{T}_+ \equiv T_+\,r_c, \,, \qquad \tilde{T}_c \equiv T_c\,r_c  \,.
\ee
 Ingoing boundary conditions at the black hole horizon requires choosing the lower sign in \eqref{NonExt:BCeventH}. 
A similar analysis around the cosmological horizon, $y=1$, yields
\be
\Phi^{-}_{\omega \ell}(y)\simeq (1-y)^{\pm\,\frac{i\,\tilde{\omega}}{2\pi \tilde{T}_c}}\big[1+\mathcal{O}(1-y)\big]\,,
\label{NonExt:BCcosmotH}
\ee
where $\tilde{T}_c$ is the (dimensionless) cosmological horizon temperature defined below \eqref{metricRNaux}. Imposing outgoing boundary conditions at the cosmological horizon demands choosing the lower sign in \eqref{NonExt:BCcosmotH}.
We will use a Chebyshev collocation scheme to numerically solve for $\left(\omega r_c,\Phi^{-}_{\omega \ell}\right)$ (see \cite{Dias:2015nua} for a review on the subject). Hence, we want to perform a field redefinition that automatically enforces the above boundary conditions. This motivates the wavefunction redefinition
\be
\Phi^{-}_{\omega \ell}(y)= y^{-\frac{i\,\tilde{\omega}}{2\pi \tilde{T}_+}}(2 - y^2)^{-\frac{i\,\tilde{\omega}}{4\pi \tilde{T}_+}}
(1-y^2)^{-\,\frac{i\,\tilde{\omega}}{2\pi \tilde{T}_c}}\mathcal{Q}_{\omega\ell}(y)\,,
\label{NonExt:redef}
\ee
where, for our choice of boundary conditions, $\mathcal{Q}_{\omega\ell}(y)$ is a smooth function of $y$ with a regular Taylor series expansion at both $y=0$ and $y=1$. The boundary conditions for $\mathcal{Q}_{\omega\ell}(y)$ are then of the Neumann type, \emph{i.e.} $\partial_y\,\mathcal{Q}_{\omega\ell}(y)\big|_{y=0,1}=0$.

Consider now the eigenvalue problem II listed above. This time we have an extremal black hole, meaning that at the event horizon $f$ vanishes quadratically. It follows that, instead of  \eqref{NonExt:BCeventH},  this time a Frobenius analysis about $y=0$  yields
\be
\Phi^{-}_{\omega \ell}(y)\simeq e^{\pm i\,\tilde{\omega} \frac{\alpha}{y^2}} y^{\pm\,i\,\tilde{\omega} \beta}\big[1+\mathcal{O}(y)\big]\,,
\label{Ext:BCeventH}
\ee
where 
{\small
\begin{eqnarray}
&& \alpha\equiv \frac{y_+^2}{2 (n-1) (1-y_+)}\,\frac{(n+1) y_+^{2 n}-2 n y_+^{n+1}+n-1}{4 n y_+^{n+1}-(n+1) \left(y_+^{2 n}+n y_+^2\right)+(n-1)^2}\,, \\
&& \beta\equiv 2 n y_+ \!\left[(n+1) y_+^{2 n}-2 n y_+^{n+1}+n-1\right]
\frac{4 (1-2 n) y_+^{n+1}+(n+1) \left[y_+^{2 n}+(3 n-2) y_+^2\right]-3 (n-1)^2}
{3 (n-1) \left[4 n y_+^{n+1}-(n+1) \left(y_+^{2 n}+n y_+^2\right)+(n-1)^2\right]^2}. \nonumber
\end{eqnarray}%
}
Requiring ingoing boundary conditions at the event horizon of the extremal RNdS black hole amounts to choose the upper sign in \eqref{Ext:BCeventH}. For the extremal RNdS, a Taylor expansion about the cosmological horizon still yields \eqref{NonExt:BCcosmotH} and choosing the lower sign in \eqref{NonExt:BCcosmotH} still amounts to impose outgoing boundary conditions at the cosmological horizon, alike in the non-extremal case. Much like in the non-extremal case, since we use a pseudospectral collocation scheme, these boundary conditions are best implemented if we introduce the field redefinition 
\be
\Phi^{-}_{\omega \ell}(y)=  e^{\frac{i\,\tilde{\omega} \alpha}{y^2(2-y^2)}} y^{i\,\tilde{\omega} \beta}
(1-y^2)^{-\frac{i\,\tilde{\omega}}{2\pi \tilde{T}_c}}\,\mathcal{Q}_{\omega\ell}(y)\,,
\label{Ext:redef}
\ee
such that the above boundary conditions simply translate into Neumann conditions, namely $\partial_y\,\mathcal{Q}_{\omega\ell}(y)\big|_{y=0,1}=0$ for the smooth function $\mathcal{Q}_{\omega\ell}(y)$.

  Finally, let us consider the nonlinear eigenvalue problem III listed above. In this case we want to find the eigenpair  $\left(q/q_{\mathrm{ext}},\Phi^{-}_{\omega \ell}\right)$ that describes the onset of the instability in the non-extremal RNdS as we scan the $y_+$ parameter. The boundary conditions for this problem are straighforward: a Frobenius analysis about $y=0$ ($y=1$) indicates that we have a term that diverges as $A \log y$ ($a \log(1-y)$). We impose boundary conditions that eliminate these divergent terms: $A\equiv 0, a\equiv 0$ by taking pure Neumann boundary conditions for $\Phi^{-}_{0\;\ell}(y)$.

\subsection{Numerical results} 
\label{sec:results2} 

The near horizon analysis of the extremal RNdS black hole and associated Durkee-Reall criterion \cite{Durkee:2010ea,Hollands:2014lra} discussed in section \ref{sec:NH} suggests that near-extremal black holes should be unstable for $n\geq 4$. To confirm this is indeed the case, we first search for unstable modes directly in the extremal RNdS black hole using the numerical scheme II outlined in the previous section \ref{sec:results1}. We find three main results: 
\begin{enumerate}

\item The gravitational instability in the RNdS black hole is present when the spacetime dimension satisfies $n\geq 4$ ($d\geq 6$) and $\ell=2$. 
  In particular, it is present for $d=6$ ($n=4$), as predicted by the near horizon criterion, which is a result that was not established in previous literature. This is explicitly shown for $n\geq 5$ in the right panel of Fig. \ref{fig:extreme} and for $n=4$ in left panel of  Fig. \ref{fig:extreme}. These plots display the imaginary part of the dimensionless frequency, ${\rm Im}(\omega r_c)$ as a function of $y_+=r_+/r_c$ (the real part of the frequency of the unstable modes vanishes). 
  \begin{figure}[ht]
  \centering 	
	\centering 	\includegraphics[width=0.46\textwidth]{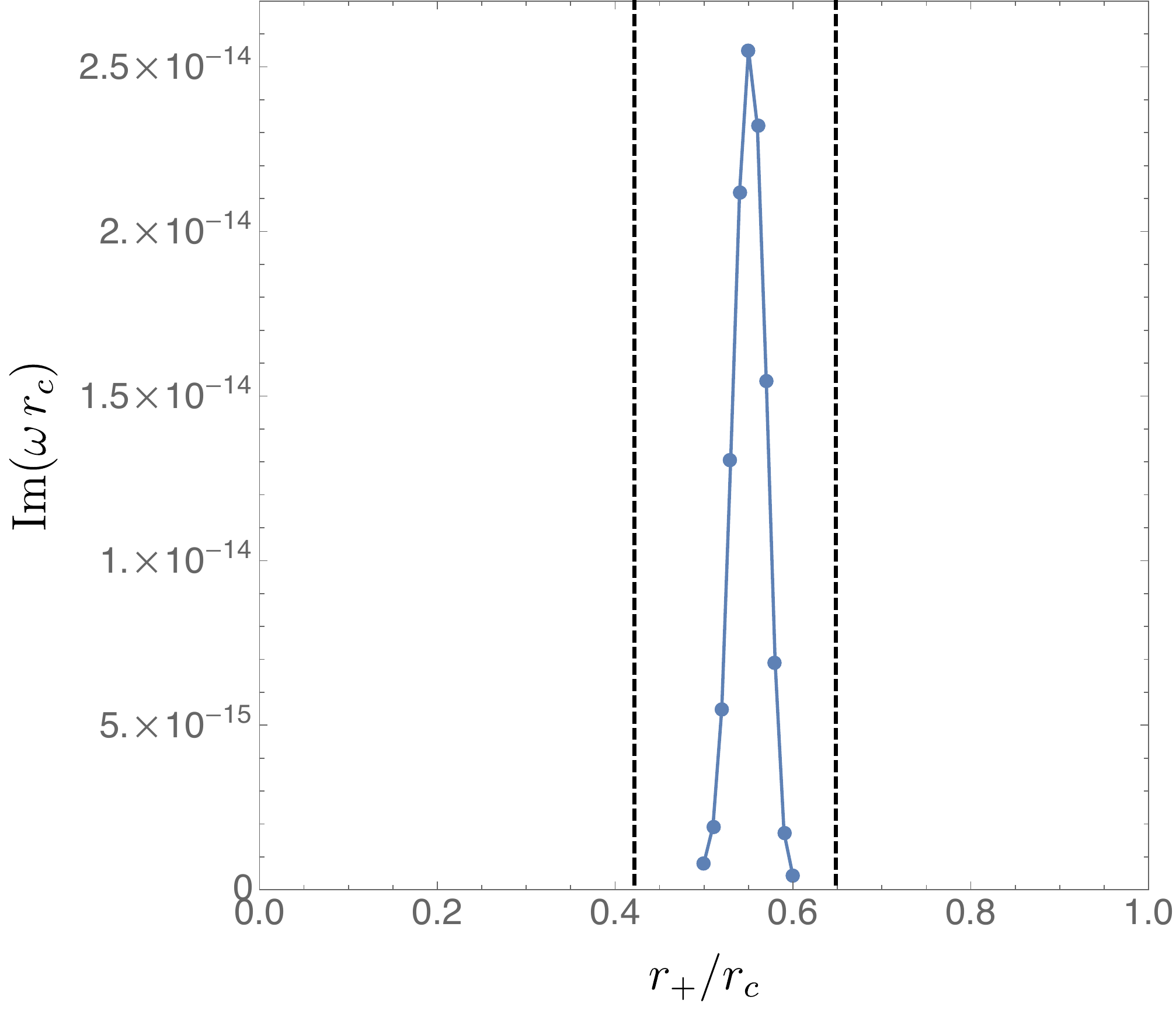} 
	\hspace{0.2cm}
	\includegraphics[width=0.46\textwidth]{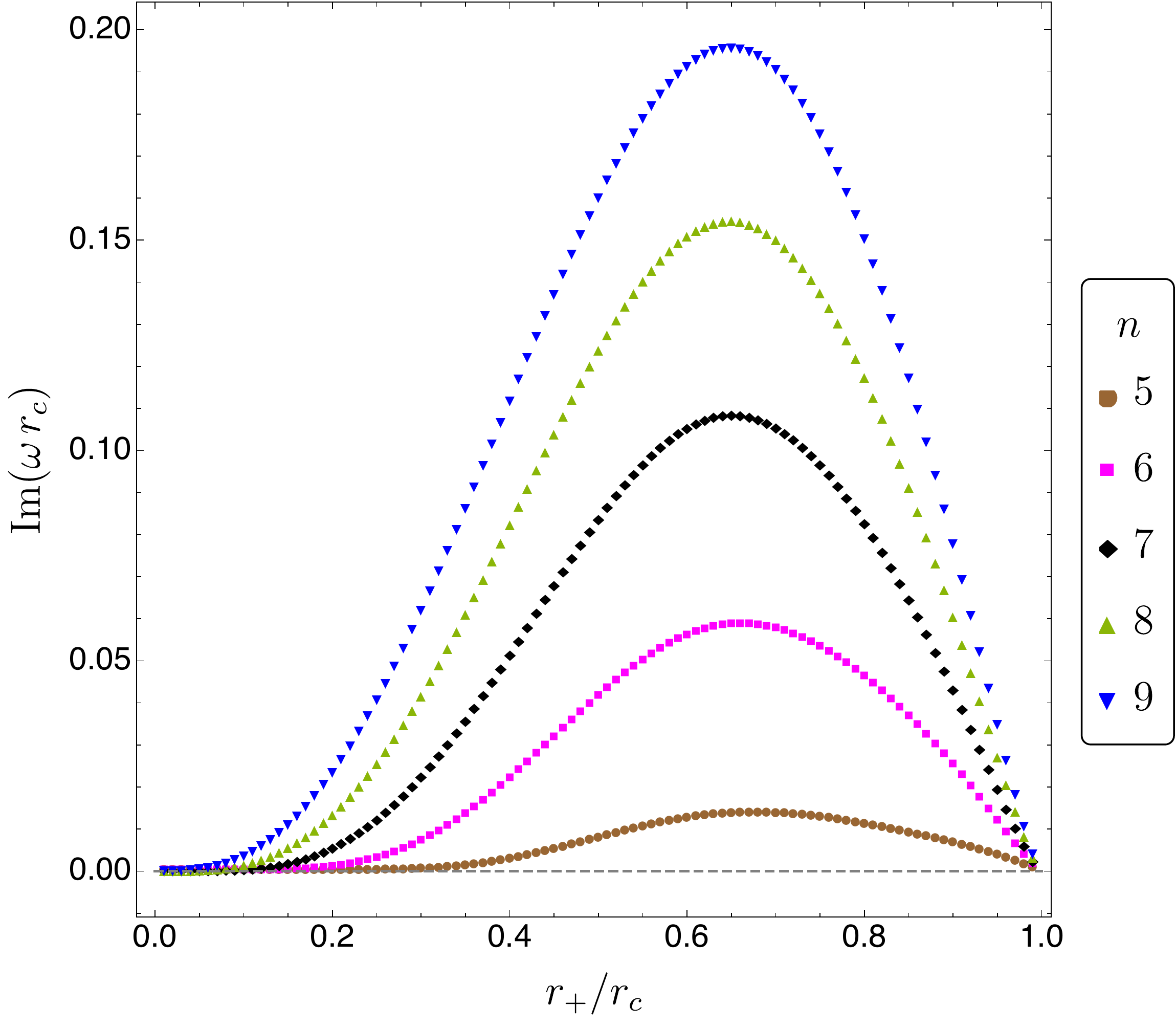} 
		\caption{Instability timescale at extremality, $q=q_{\rm ext}$ as a function of $r_+/r_c$  (the real part of the frequency of these unstable modes vanishes). 
		{\it Left panel}: The $d=6$ ($n=4$) case: the Durkee-Reall criterion is satisfied in between the two black dashed vertical lines. {\it Right Panel}: The distinct curves describe different spacetime dimensions, $d=n+2\geq 5$ (see legend). The instability is stronger for higher $n$ but, at extremality, it is always present for any value of $r_+/r_c$ in the allowed range $]0,1[$.} 
		\label{fig:extreme}
 \end{figure}

\item At extremality, $q/q_{\mathrm{ext}}=1$, the instability is present for {\it any} value of $y_+=r_+/r_c$ {\it i.e.} for $y_+\in [0,1]$ (when $n\geq 4$ and $\ell=2$) with ${\rm Im}(\omega r_c) \to 0$ as $y\to 0$ or $y\to 1$ and attaining a maximum in between. This is explicitly shown for $n\geq 5$ in the right panel of Fig. \ref{fig:extreme}. For $n=4$ it is computationally much harder to generate the associated numerical data but we believe this case should behave much like the $n\geq 5$ cases. Note that in previous literature \cite{Konoplya:2008au,Cardoso:2010rz}, the presence of the instability was established only for large values of $r_+/r_c$ (and $n\geq 5$) and later it was (incorrectly) claimed that the instability is present only above a critical value of $y_+$  \cite{Konoplya:2013sba}. 

\item The near horizon Durkee-Reall criterion \cite{Durkee:2010ea,Hollands:2014lra}  discussed in section \ref{sec:NH}  provides a sufficient but not necessary condition for the instability. That is to say, our numerical results show that the system is indeed always unstable whenever the AdS$_2$ BF bound is violated. But the instability {\it also extends} to values of the parameter space where the near-horizon criterion does not signal an instability. This is best seen comparing the analytical near-horizon predictions of Fig. \ref{fig:NH} with the actual numerical results of Fig. \ref{fig:extreme}. Recall that both these plots are for $\ell=2$ and $q=q_{\rm ext}$. Typically,  the near-horizon analysis of Fig. \ref{fig:NH} predicts instability for a {\it finite} range of $y_+$. However, we find that, at extremality, the system is unstable in the {\it full} range  $y_+\in [0,1]$. This is certainly the case for $n \geq 5$ (see right panel of Fig. \ref{fig:extreme}) and should also be true for $n=4$. 

\end{enumerate}
 
\begin{figure}[t]
	\centering 	\includegraphics[width=0.46\textwidth]{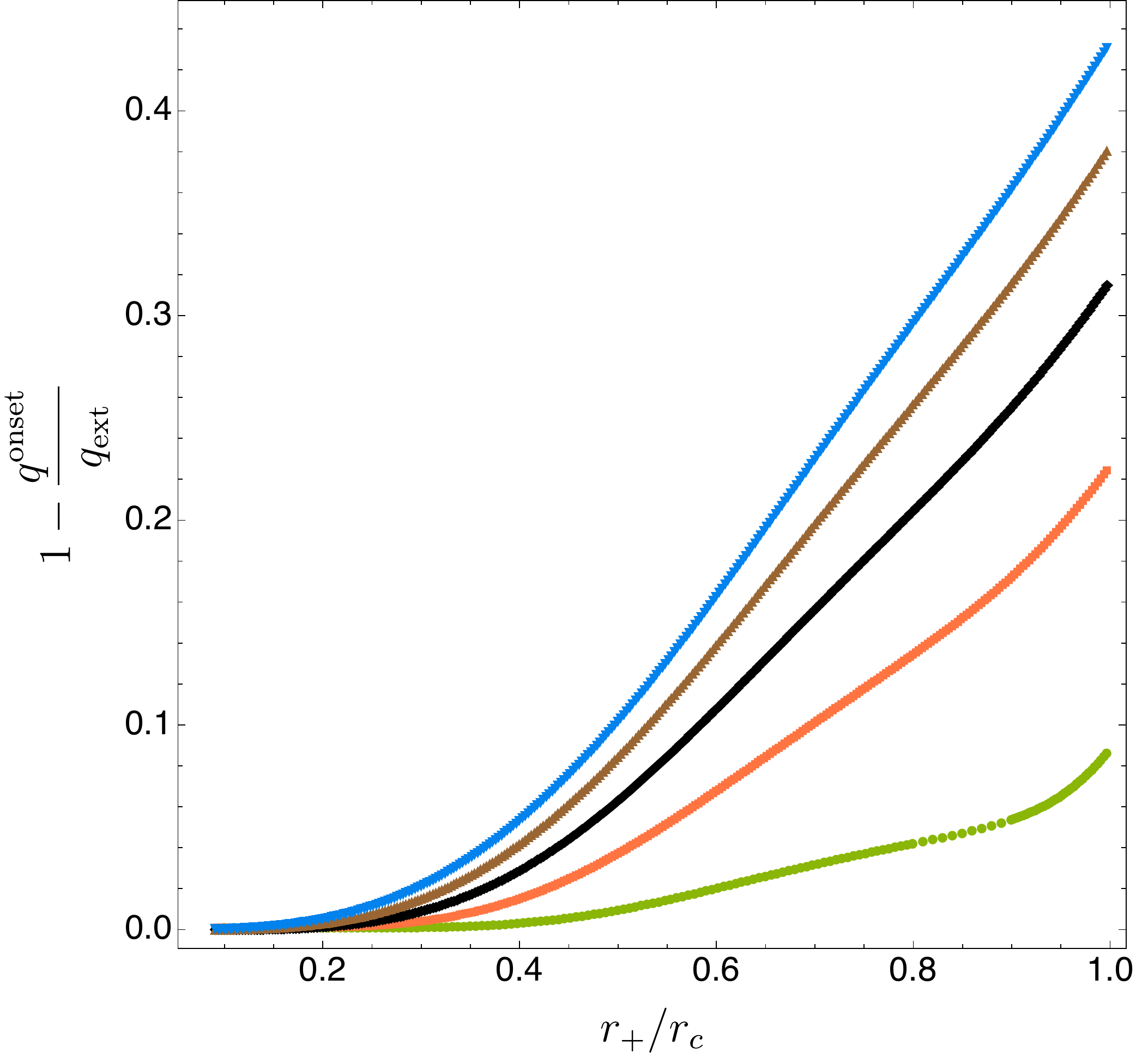} \hspace{0.1cm}
	\centering 	\includegraphics[width=0.51\textwidth]{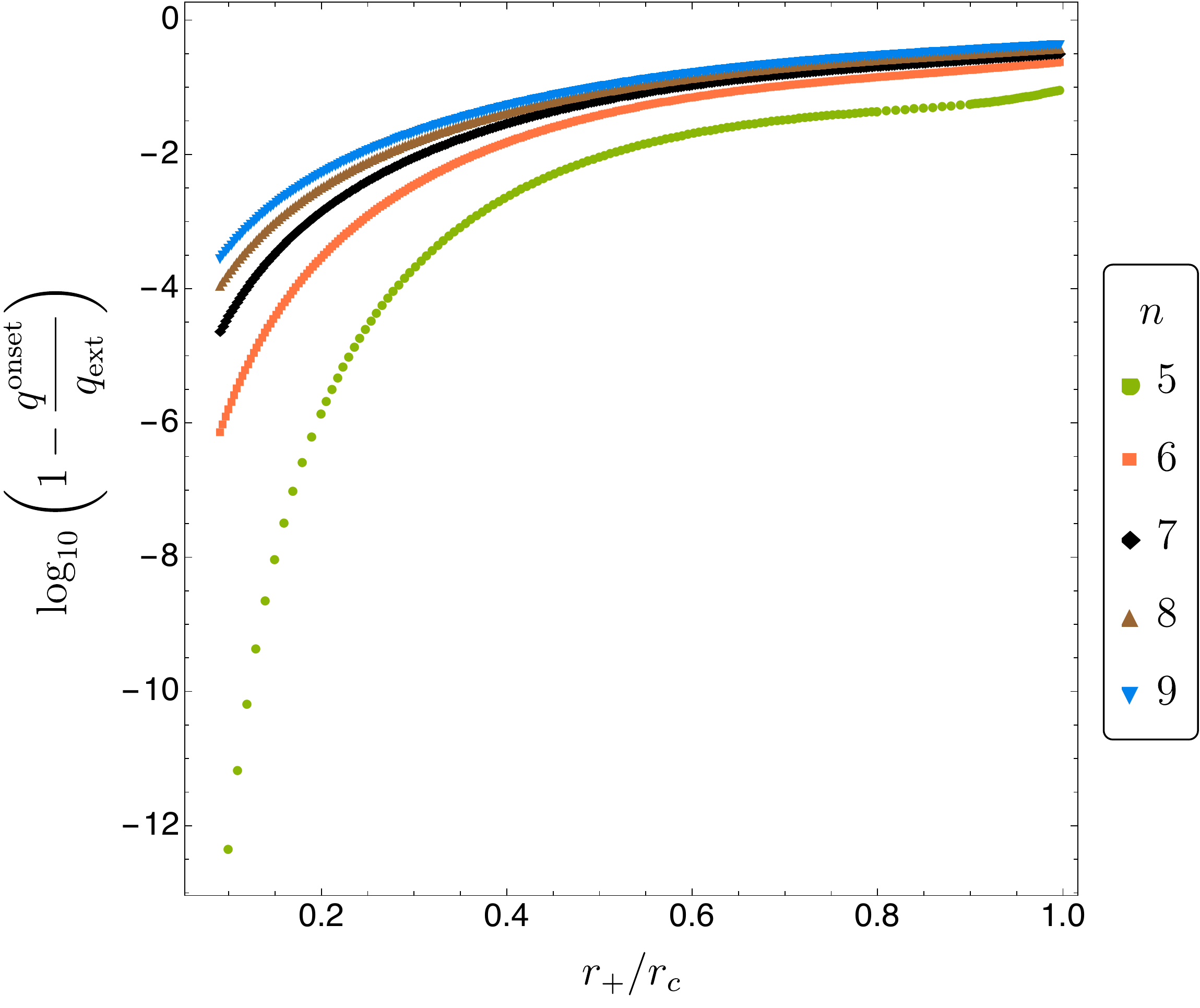} 
		\caption{Instability onset. For given $r_+/r_c$, RNdS black holes with $q>q^{\rm onset}$ are unstable. The right panel is simply a log plot of the left panel to better see what happens for small $y_+$.} 
		\label{fig:onset}
 \end{figure}

 Having established that all, $0 \leq y_+\leq 1$, extremal RNdS black holes are  unstable for $n\geq 4$ we might then ask how far away from extremality does the instability extend into. To address this  
 question we first search directly for the {\it onset} of the instability using the numerical scheme III outlined in the previous section \ref{sec:results1}. This critical charge $q^{\mathrm{onset}}$ above which the RNdS solution is unstable is shown in Fig. \ref{fig:onset} for $n=5,6,7,8,9$ and  $\ell=2$. The left panel shows $1-q^{\mathrm{onset}}/q_{\mathrm{ext}}$ as a function of $y_+$ while the right panel shows the logarithmic plot of the same quantity to zoom the details of the small $y_+$ region. We see that  for a given dimension $n$, $q^{\mathrm{onset}}/q_{\mathrm{ext}}$ decreases as $y_+$ grows from 0 into 1: the instability extends further away from extremality for high $y_+$. On the other hand, for a given $y_+$ we see that increasing the dimension $n$ favours the instability in the sense that  $q^{\mathrm{onset}}/q_{\mathrm{ext}}$ becomes smaller as $n$ grows.
 
  \begin{figure}[bh]
	\centering 	\includegraphics[width=0.6\textwidth]{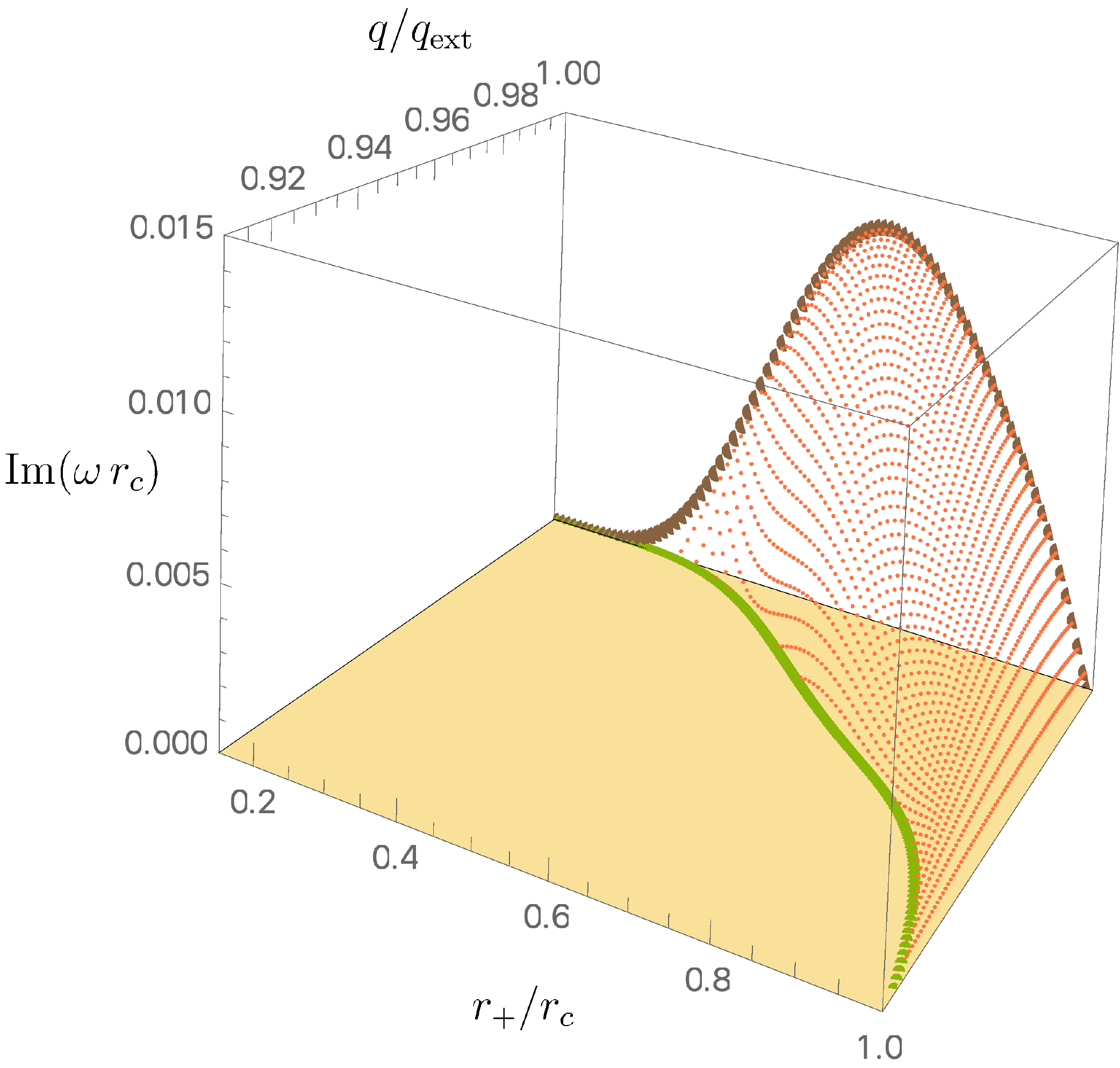} 
		\caption{Instability timescale for $n=5$: imaginary part of the frequency ${\rm Im}(\omega r_c)$ as a function of $r_+/r_c$ and $q/q_{\rm ext}$ (only shown the parameter space region where the instability is present, {\it i.e.} that has  ${\rm Im}(\omega r_c)\geq 0$). The brown dots in the plane $q/q_{\rm ext}=1$ represent instability data collected independently using a numerical code at extremality. The green dots with ${\rm Im}(\omega r_c)=0$ represent data collected independently using a numerical code for the onset of the instability.} 
		\label{fig:3D}
 \end{figure} 
 
\begin{figure}[ht]
	\centering 	\includegraphics[width=0.47\textwidth]{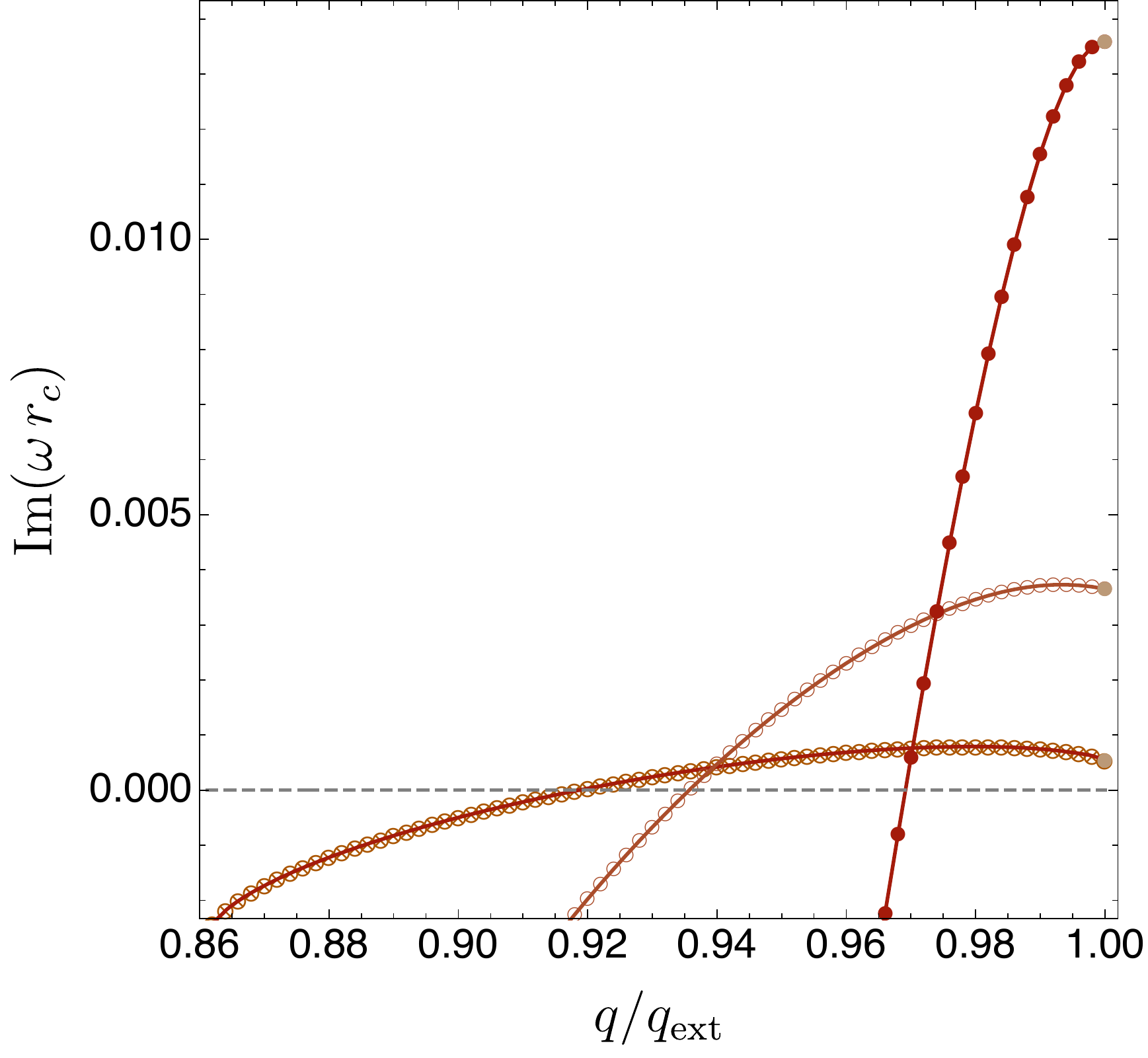}
		\caption{Frequency spectrum (imaginary part) for RNdS black hole families with $y_+=0.70$ ($\bullet$),  $y_+=0.95$  ($\circ$) and $y_+=0.99$  ($\otimes$) and $n=5$, $\ell=2$. The brown disks with $q/q_{\rm ext}=1$were obtained using the independent code II for the extremal solution. The onset of the instability (where ${\rm Im}(\omega r_c)=0$) occurs at the critical values $\{y_+,q/q_{\rm ext}\}$ also obtained using the independent numerical code III used to generate the onset curve of Fig. \ref{fig:onset}.} 
		\label{fig:severalyp}
 \end{figure}

 To have a broader perspective of the properties of the instability, next we search directly for the {\it instability timescale} in the {\it non}-extremal RNdS black hole. For this discussion we fix the dimension to be $n=5$ and the associated data is shown in Fig. \ref{fig:3D} (for other dimensions the plot is qualitatively similar).  Recall that non-extremal RNdS black holes are parametrized by $q/q_{\mathrm{ext}}$ and $r_+/r_c$ and these are the two horizontal axes of Fig. \ref{fig:3D}. On the other hand the vertical axis is the imaginary part of the dimensionless frequency, ${\rm Im}(\omega r_c)$ (the real part of the frequency of the unstable modes vanishes). In this 3-dimensional plot we also show the extremal (brown) curve already displayed in Fig. \ref{fig:extreme} and the onset (green) curve already shown in Fig. \ref{fig:onset} (for $n=5$). The fact that the 2-dimensional surface describing the regime where RNdS is unstable  ends on the extremal and onset curves obtained using independent numerical codes provides a non-trivial check of our numerical results. This plot reveals the following properties (some were already discussed above): 
 \begin{itemize}
 \item At extremality, $q/q_{\mathrm{ext}}=1$, the instability is present in the whole range $0 \leq y_+\leq 1$.
 \item As we move away from extremality, we see that for a given (large) charge, namely in the range $0.9128 \lesssim q/q_{\rm ext}\leq 1$ (when $n=5$) the system is unstable only for $y_+$ above a (non-vanishing) critical value. For smaller charges $q$ the system is stable. In equivalent words, for a given $y_+$, RNdS black holes are unstable if their charge is above $q^{\mathrm{onset}}$, with $q^{\mathrm{onset}}$ approaching $q_{\mathrm{ext}}$ as $y_+\to 0$.
 
 \item The maximum strength of the instability is attained for black holes that are close to extremality, but not at extremality. This is better seen in Fig. \ref{fig:severalyp} where we show the instability timescale for three families of RNdS solutions at constant $y_+$ (namely $y_+=0.70, 0.95, 0.99$) as a function of the dimensionless charge ratio $q/q_{\mathrm{ext}}$. We see that typically  ${\rm Im}(\omega r_c)$ grows as   $q/q_{\rm ext}$ increases but its maximum occurs slight before one reaches the extremal configuration   $q/q_{\rm ext}=1$.
   \end{itemize}

\begin{figure}[th]
	\centering 	\includegraphics[width=0.6\textwidth]{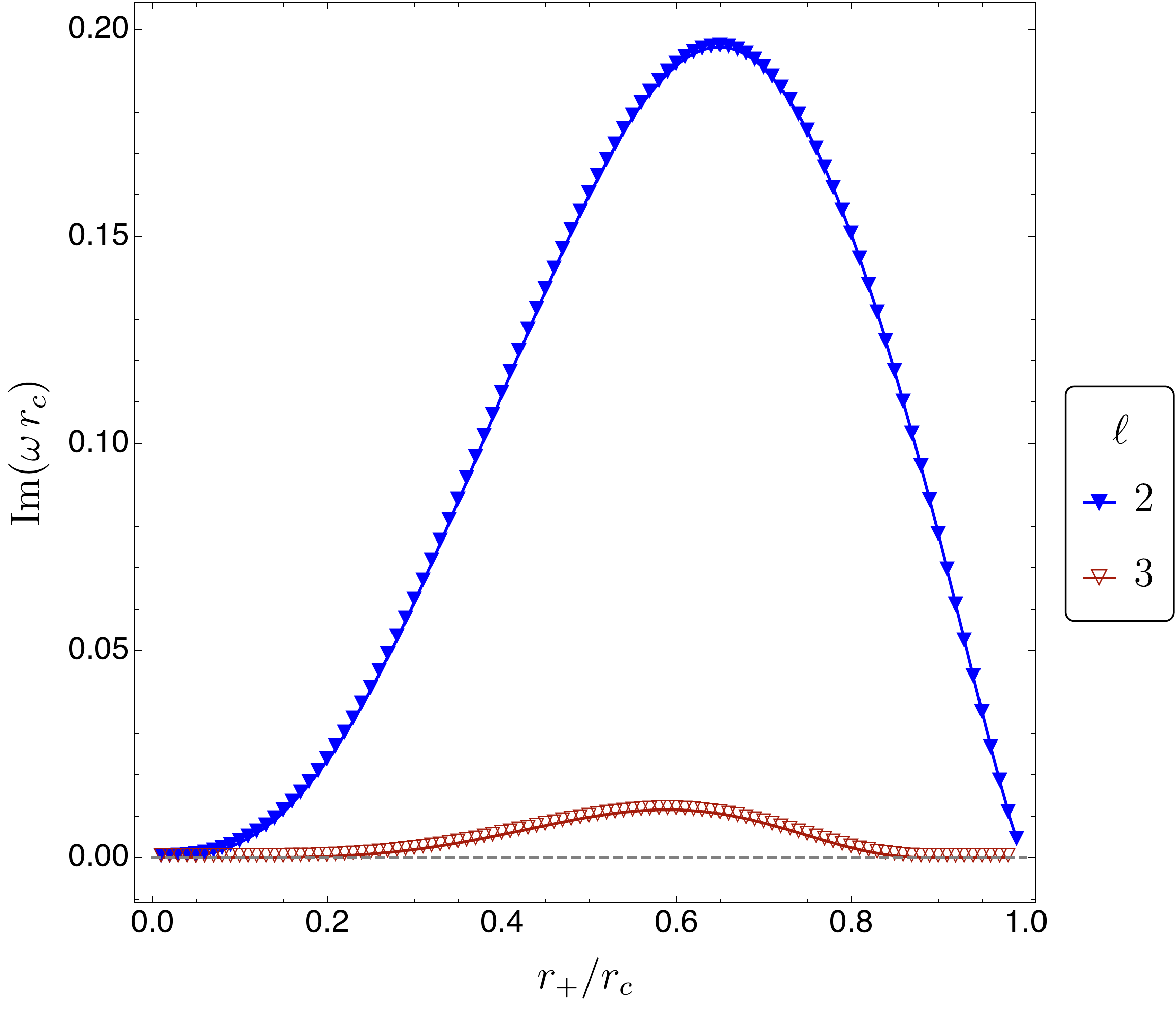} 
		\caption{Instability timescale  at extremality, $q=q_{\rm ext}$, for $n=9$ as a function of $r_+/r_c$ for the $\ell=2$ (blue $\blacktriangledown$) and $\ell=3$ (red $\triangledown$) harmonics. The $\ell=2$ curve was already shown (with same colour/shape code) in Fig. \ref{fig:extreme}.} 
		\label{fig:harmonic}
 \end{figure}

Up until now we have discussed only modes with $\ell=2$. This is because for each dimension $n\geq 4$ this is either: i) the only harmonic for which the instability is present or ii) the harmonic mode where the instability is stronger. But we can also discuss briefly other harmonics. 
As discussed in section \ref{sec:NH}, the near-horizon analysis shows that as $\ell$ increases it becomes harder to get negative  $\left(\mu^2-\mu^2_{\rm BF}\right) L^2_{2}$. For example, in $n=4$ or $n=5$ dimensions, for $\ell\geq 3$ the AdS$_2$ BF bound is no longer violated for any $y_+$. As another example, for $n=6,7$ the harmonics $\ell=2,3$  generate a violation of the AdS$_2$ BF bound but this is no longer the case for $\ell\geq 4$. As a final example,  for $n=8,9$ the harmonics $\ell=2,3,4$  generate a violation of the AdS$_2$ BF bound but this is no longer the case for $\ell\geq 5$. The full numerical analysis confirms these analytical predictions to be correct. In particular, the numerical analysis also concludes that when the instability is present for more than the $\ell=2$ harmonic, modes with lower $\ell$ are more unstable. To illustrate this, in Fig. \ref{fig:harmonic} we take an extremal RNdS BH in $n=9$ and  compare the instability timescale of $\ell=2$ and $\ell=3$ modes. We see that the $\ell=3$ instability timescale is typically two orders of magnitude smaller than the $\ell=2$ timescale. We have also explicitly checked that for $n=9$ the $\ell=4$ harmonic (but not higher $\ell$'s) is also unstable but with a strength that is $\sim 5$ orders of magnitude smaller than the $\ell=2$ instability. For example, for $y_+=1/2$ the timescale of three harmonics are:
 \be
 \ell=2: \:\:\omega r_c \simeq  1.60\times 10^{-1}\, i \,, \qquad \ell=3:  \:\: \omega r_c \simeq  9.78\times 10^{-3}\, i \,, \qquad \ell=4:  \:\: \omega r_c \simeq  1.37\times 10^{-6}\, i \,.
 \ee
 
So far we have focused our attention on describing key properties of the instability present in RNdS black holes and confirming that we can understand its origin as being due to the violation of a relevant AdS$_2$ BF bound discussed in section \ref{sec:NH}.
The relevant modes become unstable close to extremality. 

Next, we give a broader view of the properties of these perturbations and follow the unstable modes as we move away from extremality all the way down to the Schwarzschild-dS black hole with $q=0$. This analysis turns out to reveal an interesting quasinormal mode structure with properties that, to the best of our knowledge, no other known black hole quasinormal mode spectrum exhibits. In our scan of solutions we have found three main regimes  that are distinct from each other and thus deserve a separate discussion. Illustrative examples of each of these three cases
are the RNdS black holes with $y_+=0.70$,  $y_+=0.95$ and $y_+=0.99$ (and $n=5$, $\ell=2$). The main  instability properties near extremality of these three cases  were already discussed in  Fig. \ref{fig:severalyp}. Next we extend the analysis of these modes all the way towards $q=0$, {\it i.e.} well away from the region where the solutions are unstable and we also discuss ``secondary" (less unstable or stable) modes that are nevertheless relevant to have a good overview of the quasinormal mode structure. The key properties of these three families can be summarized as follows:
 
 \begin{enumerate}
 \item
The case $y_+=0.99$  represents what typically happens for RNdS black holes that have very large $y_+$ ({\it i.e.} close to unit). The relevant frequency spectrum for this case is displayed in 
 Fig. \ref{fig:yp099Im} (imaginary part)  and  Fig. \ref{fig:yp099Re} (real part of the frequency). In Fig. \ref{fig:yp099Im} we identify the dark red disk ($\bullet$) curve  that describes the (most) unstable mode that was already identified in Fig. \ref{fig:severalyp} with red $\otimes$ (and in Fig. \ref{fig:3D}) and that has ${\rm Re}(\omega r_c)=0$. In particular, the brown disk with  $q/q_{\rm ext}=1$ in this curve describes the extremal solution also identified in Fig. \ref{fig:extreme}. We see that this curve has a maximum close to extremality (as discussed previously) and, decreasing $q/q_{\rm ext}$ we find that the mode first becomes stable at $q/q_{\mathrm{ext}}\sim 0.92$ (consistent with Fig. \ref{fig:onset}) and then reaches a bifurcation point $A$ at $q/q_{\rm ext}\sim 0.861$ where it joins the light blue lozenge ($\lozenge$) and blue diamond ($\blacklozenge$) curves. As seen in the inset plot of the left panel, the light blue  $\lozenge$ branch extends all the way from $A$ down to $q=0$. As shown in the companion  Fig. \ref{fig:yp099Re}, the real part of the frequency of this branch is always non-vanishing (${\rm Re}(\omega r_c)\neq 0$) and it decreases monotonically as $q$ increases from zero towards the bifurcation point $A$ with $q/q_{\rm ext}\sim 0.861$. At this point $A$, the light blue  $\lozenge$ branch bifurcates into the dark red  $\bullet$ curve (upper branch) already discussed above and into the blue  $\blacklozenge$ curve (lower branch). These two branches ($\bullet$ and $\blacklozenge$) both have ${\rm Re}(\omega r_c)=0$. That is to say, a complex eigenvalue\footnote{Since RNdS is a static background with Killing vector $\partial_t$, if $\omega=\omega_R+i\,\omega_I$ is an eigenvalue then  $-\omega^{\ast}=-\omega_R+i\,\omega_I$ is also an eigenvalue of the frequency spectrum.}  bifurcates into two branches that have purely imaginary eigenvalues. Now let us follow the blue diamond $\blacklozenge$ branch. We see that it extends from point $A$ with $q/q_{\rm ext}\sim 0.861$ up to point $B$ with $q/q_{\rm ext}\sim 0.902$.  At this point $B$ it merges with the brown inverted triangle $\blacktriangledown$ curve (that also approaches point $B$ but coming from point $D$) and the solution  now extends for higher $q/q_{\rm ext}$ along the light blue  $\lozenge$ branch that again has ${\rm Re}(\omega r_c)\neq 0$ (see the half-circle $BC$ section of Fig. \ref{fig:yp099Re}) up to point $C$ with $q/q_{\rm ext}\sim 0.934$. At this point $C$ we have again a new bifurcation into two branches: the upper branch with orange triangles ($\blacktriangle$) and the lower branch with blue diamonds ($\blacklozenge$). Both these branches then extend independently towards extremality ($q/q_{\rm ext}=1$) in such a way that ${\rm Im}(\omega r_c)$ grows monotonically from negative into positive values (moreover ${\rm Re}(\omega r_c)= 0$ for both branches). This is best seen in the zoomed right panel of Fig. \ref{fig:yp099Im}. For completeness we should also mention the turning point $D$ seen in the left panel of  Fig. \ref{fig:yp099Im} where the brown inverted triangle $\blacktriangledown$ curve continues as the green square ($\blacksquare$). Both curves have ${\rm Re}(\omega r_c)= 0$. The green square curve extends away from $D$ to higher $q/q_{\rm ext}$ where it has a new turning point but we do not analyse/discuss this further.
   \begin{figure}[t]
	\centering 	\includegraphics[width=0.47\textwidth]{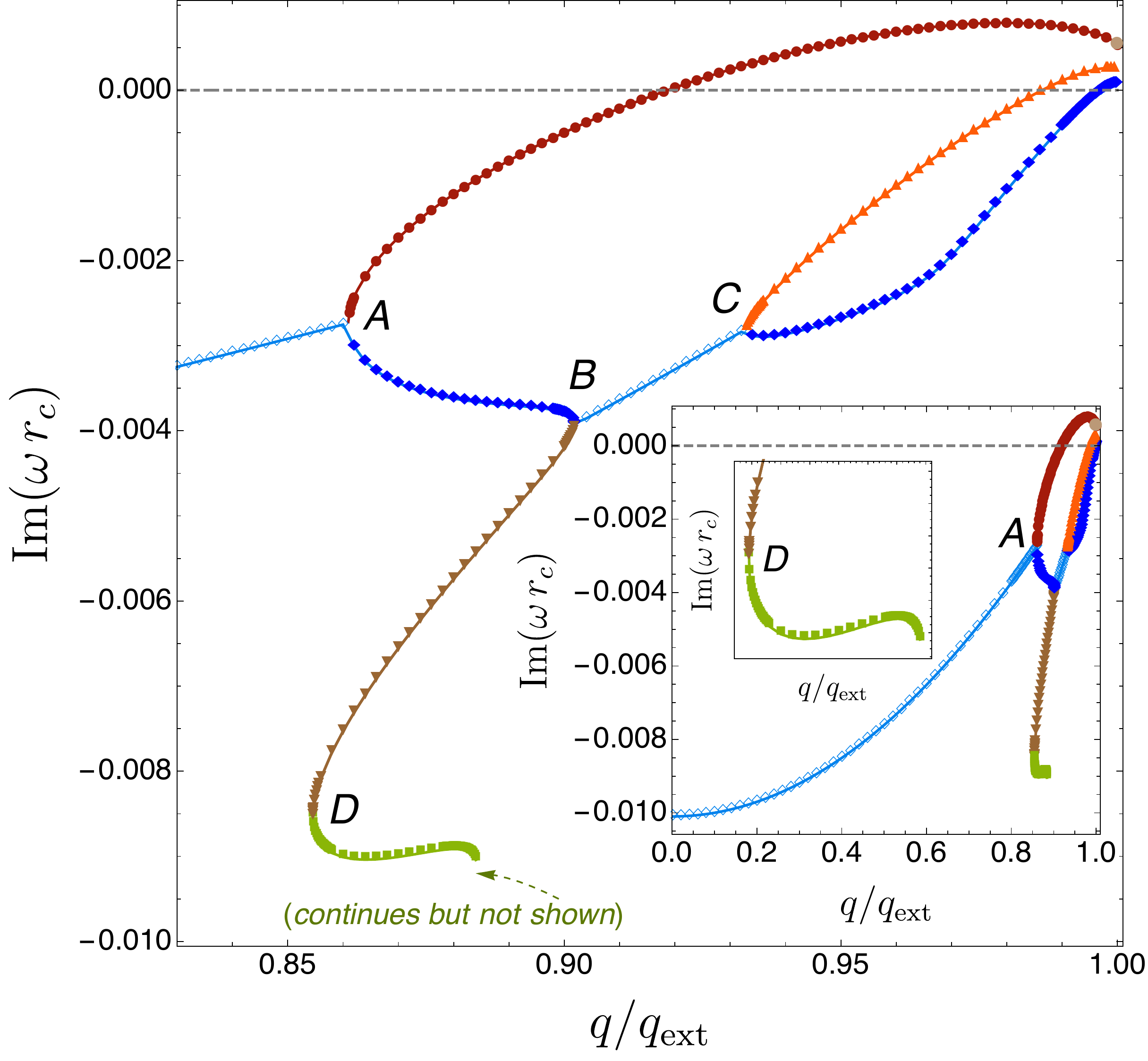} \hspace{0.3cm}
	\centering 	\includegraphics[width=0.49\textwidth]{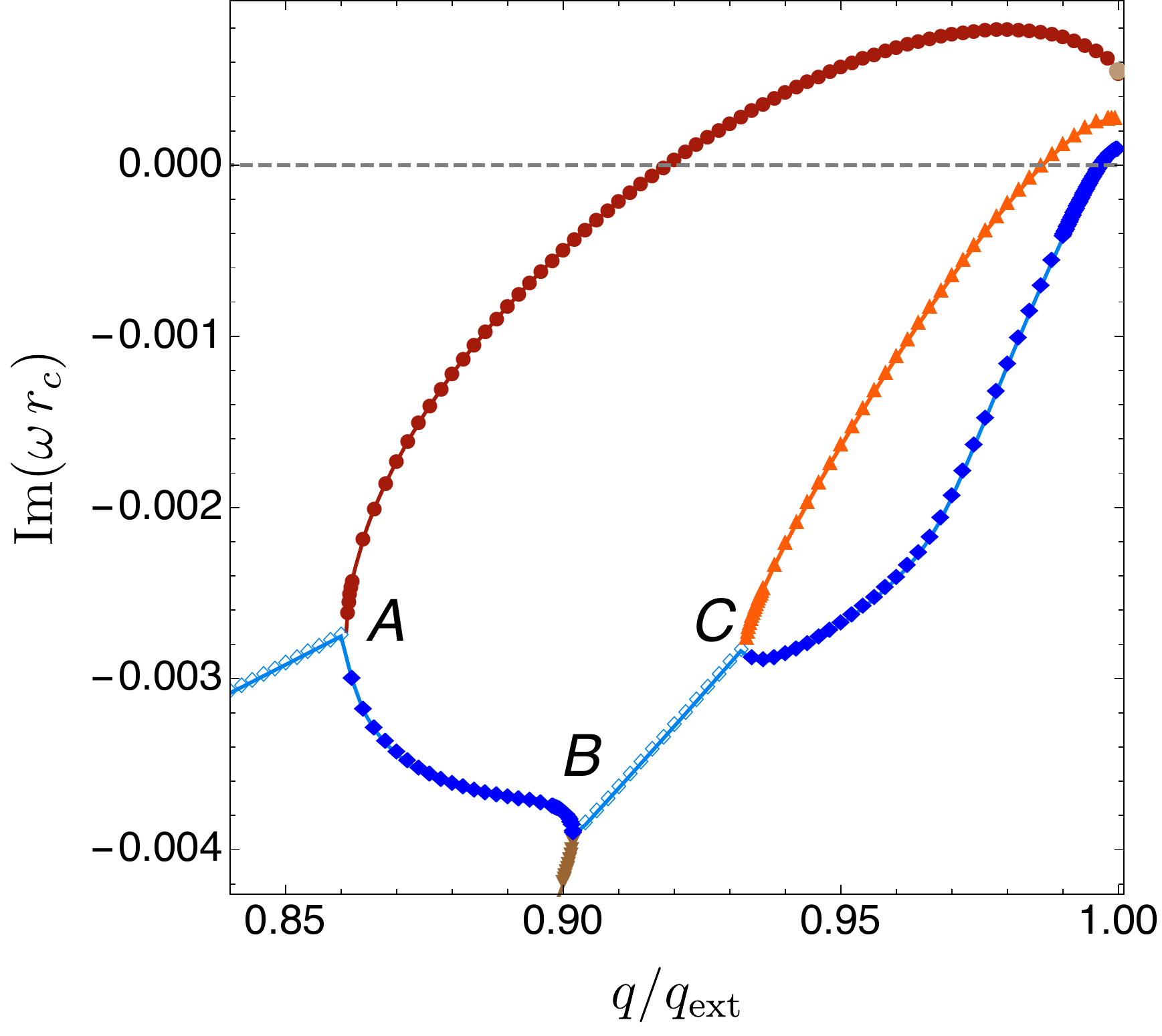} 
		\caption{Frequency spectrum (imaginary part) for $y_+=0.99$ and $n=5$, $\ell=2$ with zooms in the relevant regions where instability is present.} 
		\label{fig:yp099Im}
 \end{figure} 
    \begin{figure}[t]
	\centering 	\includegraphics[width=0.47\textwidth]{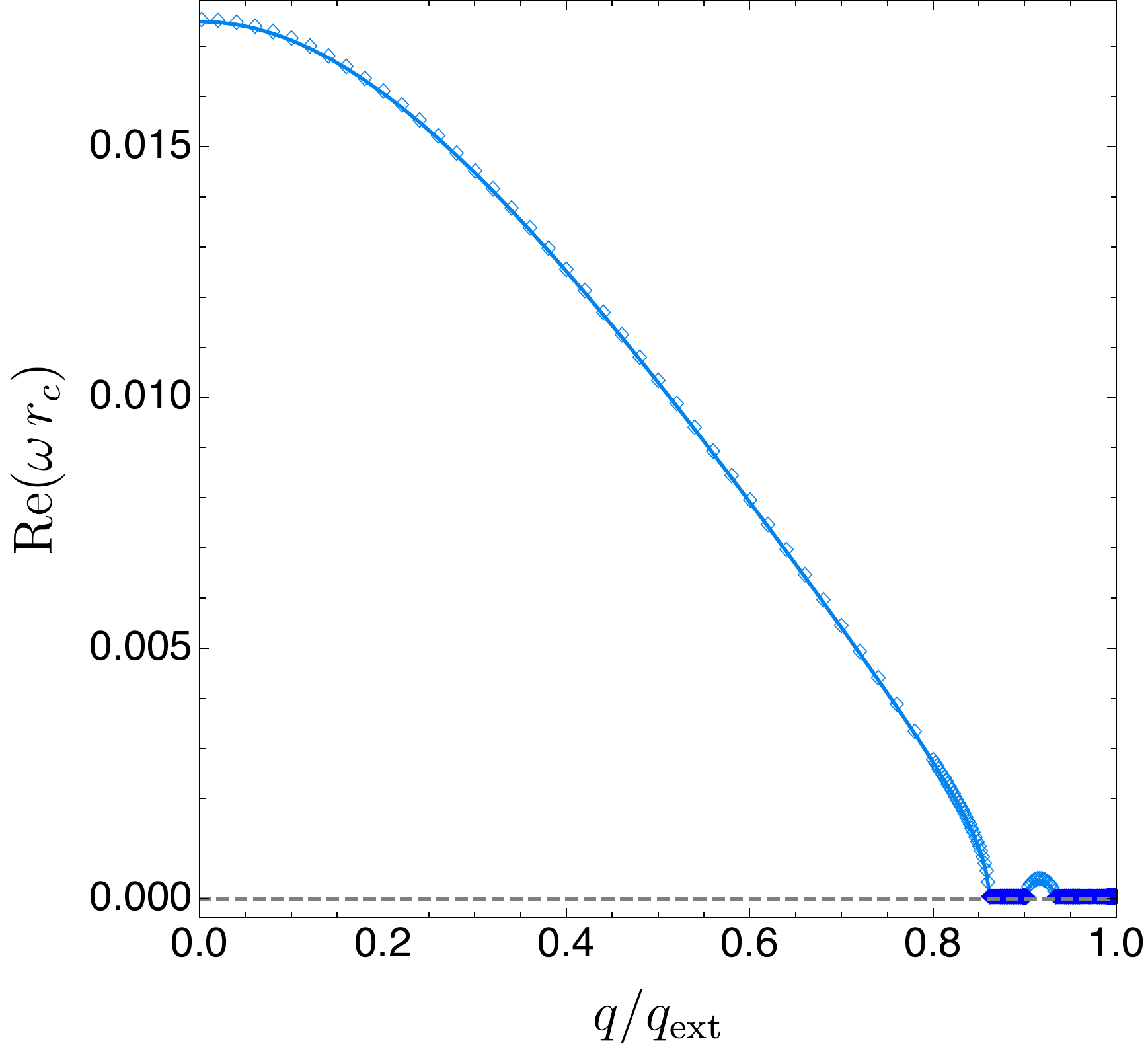} \hspace{0.3cm}
	\centering 	\includegraphics[width=0.49\textwidth]{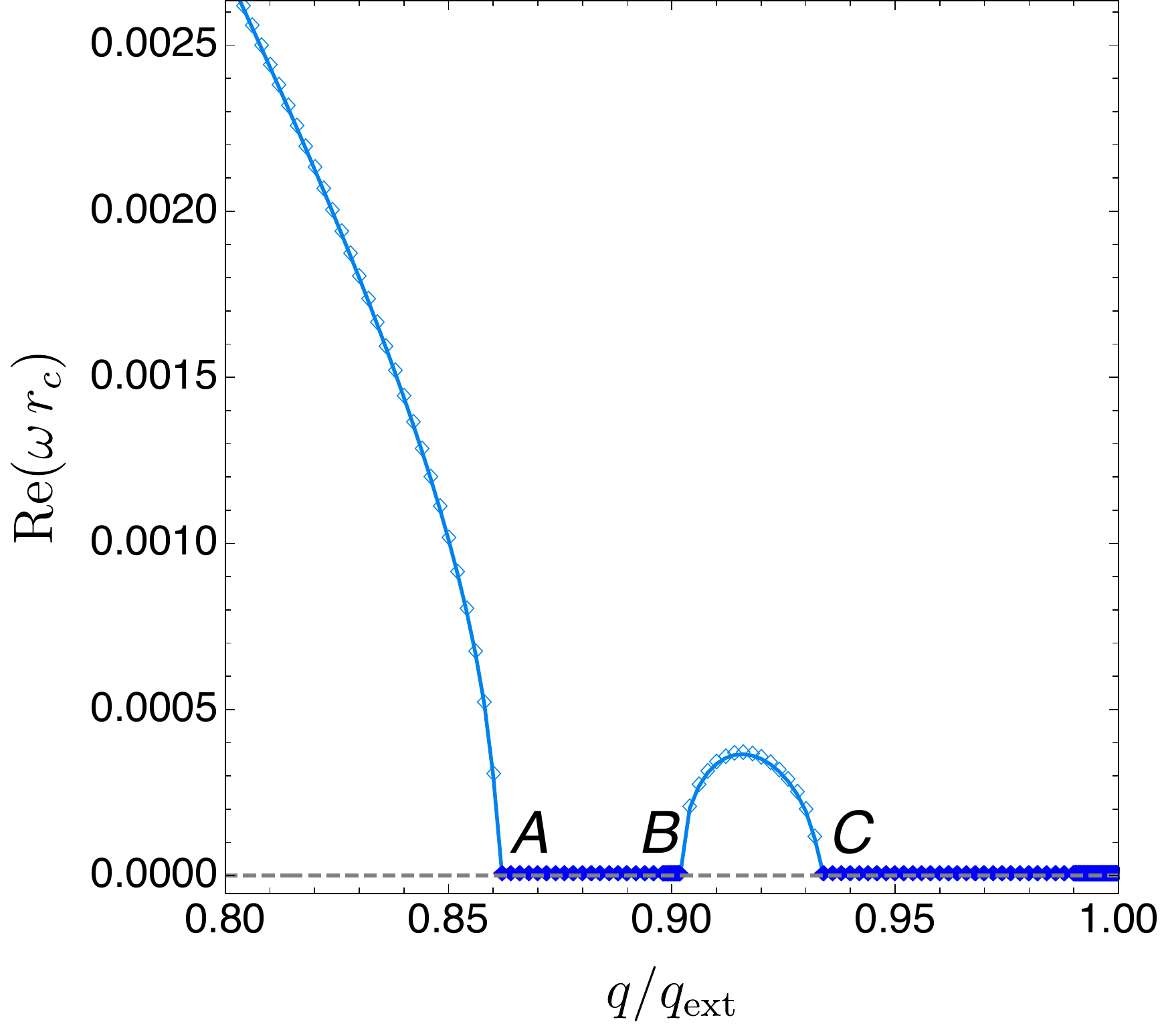} 
		\caption{Frequency spectrum  (real part) for $y_+=0.99$ and $n=5$, $\ell=2$. The right panel  zooms in the near-extremal region where the instability is present. The green, red, dark red and orange sections of Fig. \ref{fig:yp099Im} are {\it not} shown: they all have ${\rm Re}(\omega r_c)=0$.} 
		\label{fig:yp099Re}
 \end{figure} 
 
Although we have not done an exhaustive scan of the parameter space, we have collected enough evidence that the frequency spectrum for RNdS black holes with $y_+$ close to 1 has a bifurcation/merger structure qualitatively similar to the one shown in Fig. \ref{fig:yp099Im} and  Fig. \ref{fig:yp099Re} (some of this evidence comes from Figs. \ref{fig:q095Im}-\ref{fig:q095Re}, \ref{fig:q088Im}-\ref{fig:q088Re} and \ref{fig:q085Im}-\ref{fig:q085Re} to be discussed later).
It is also important to point out that a bifurcation structure similar to the one seen around point $A$ of Figs. \ref{fig:yp099Im}-\ref{fig:yp099Re} also emerges from a large $d$ analysis of the instability: see Fig. 1 of \cite{Tanabe:2015isb}. However, at least for finite values of the dimension, the system seems to have a more complicated bifurcation structure than the one reported in \cite{Tanabe:2015isb}: there are further bifurcations/mergers like $B$ and $C$ in Figs. \ref{fig:yp099Im}-\ref{fig:yp099Re}.  
 
  \item Next we consider the case $y_+=0.95$  which represents RNdS that have moderately large $y_+$ that is however not large enough to be close to unit. The relevant frequency spectrum for this case is displayed in  Fig. \ref{fig:yp095Im} (imaginary part)  and  Fig. \ref{fig:yp095Re} (real part of the frequency). 
     \begin{figure}[b]
	\centering 	\includegraphics[width=0.47\textwidth]{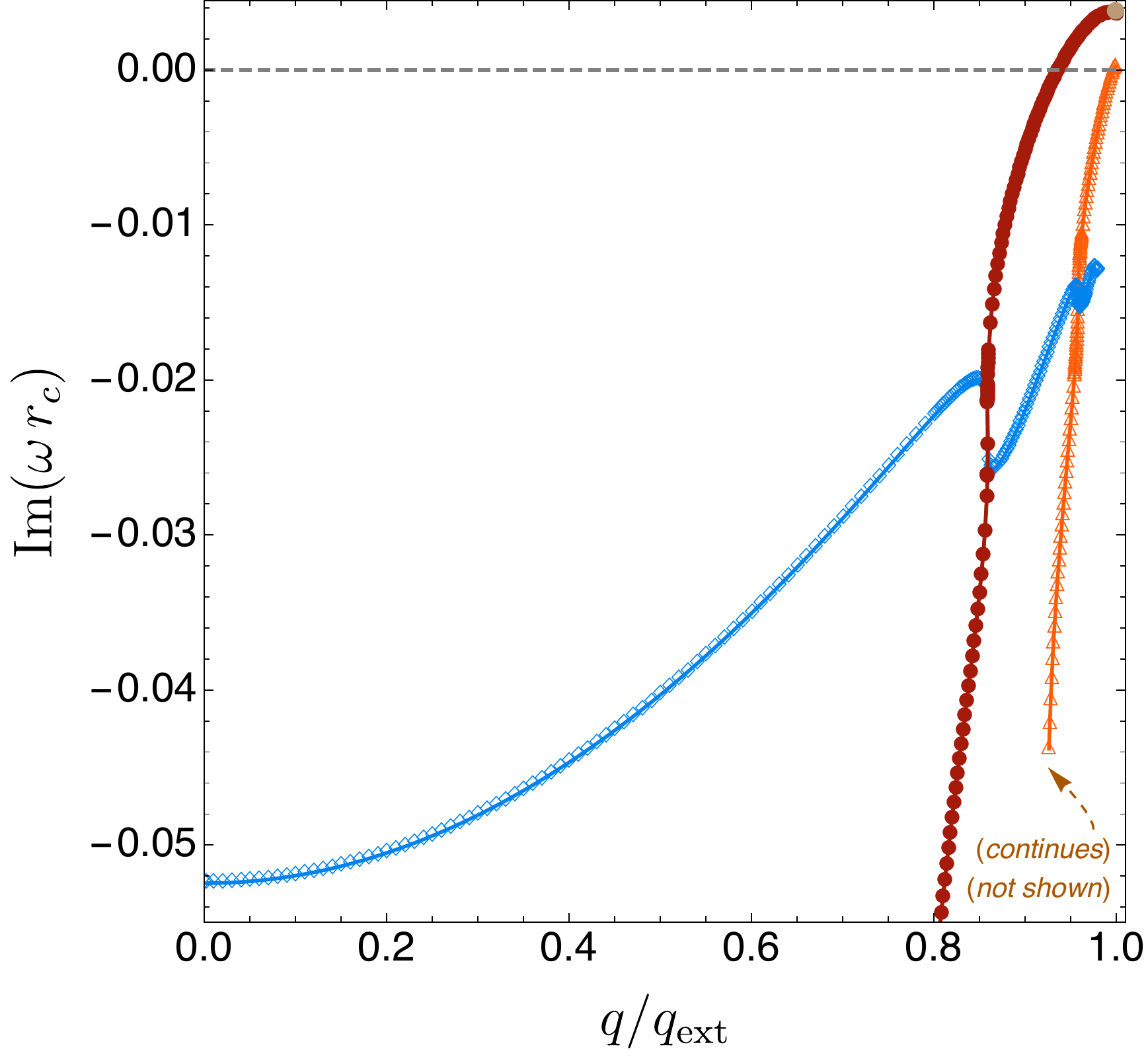} \hspace{0.3cm}
	\centering 	\includegraphics[width=0.475\textwidth]{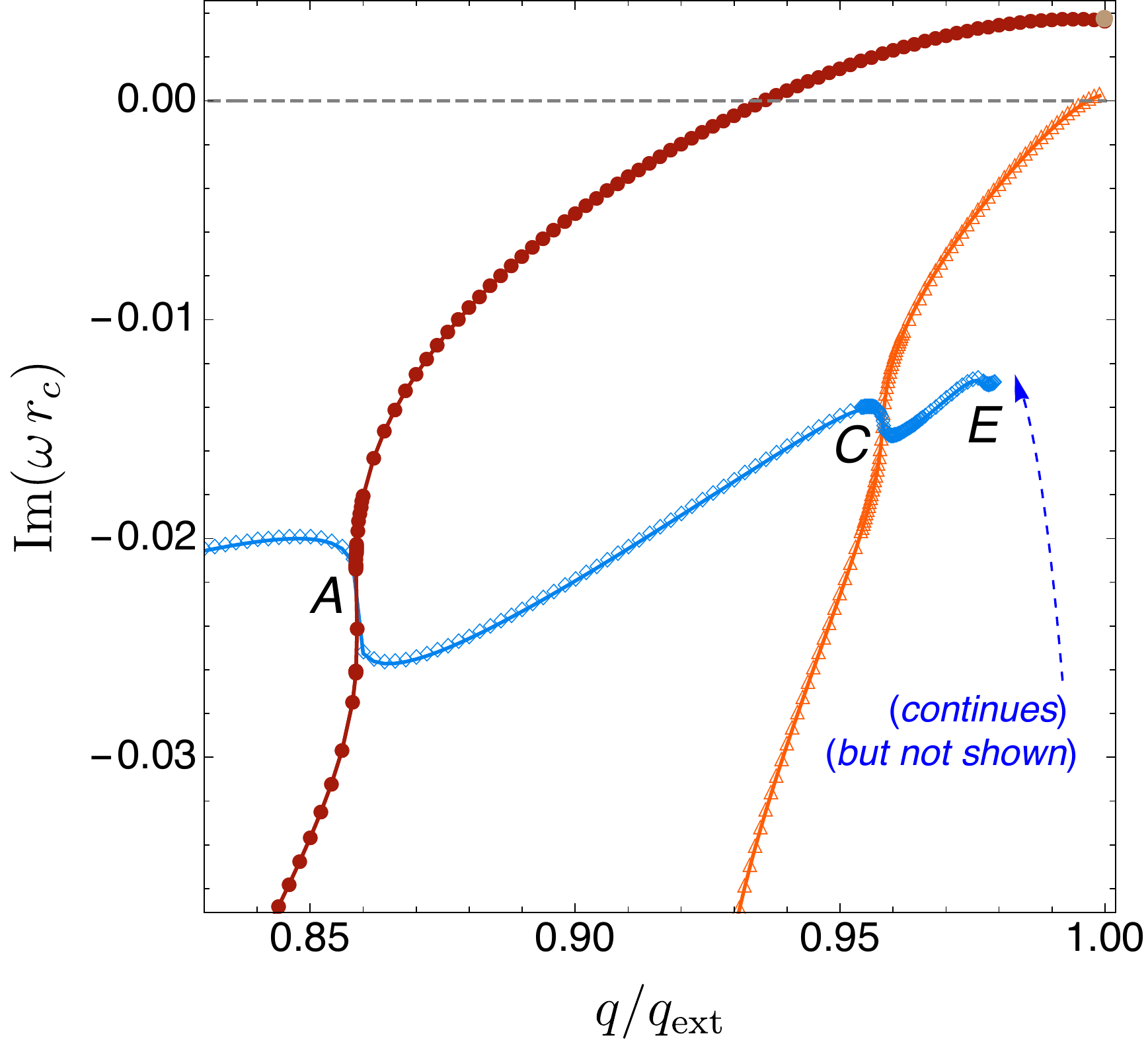} 
		\caption{Frequency spectrum (imaginary part) for $y_+=0.95$ and $n=5$, $\ell=2$ with zooms in the relevant regions where instability is present.} 
		\label{fig:yp095Im}
 \end{figure} 
     \begin{figure}[th]
	\centering 	\includegraphics[width=0.47\textwidth]{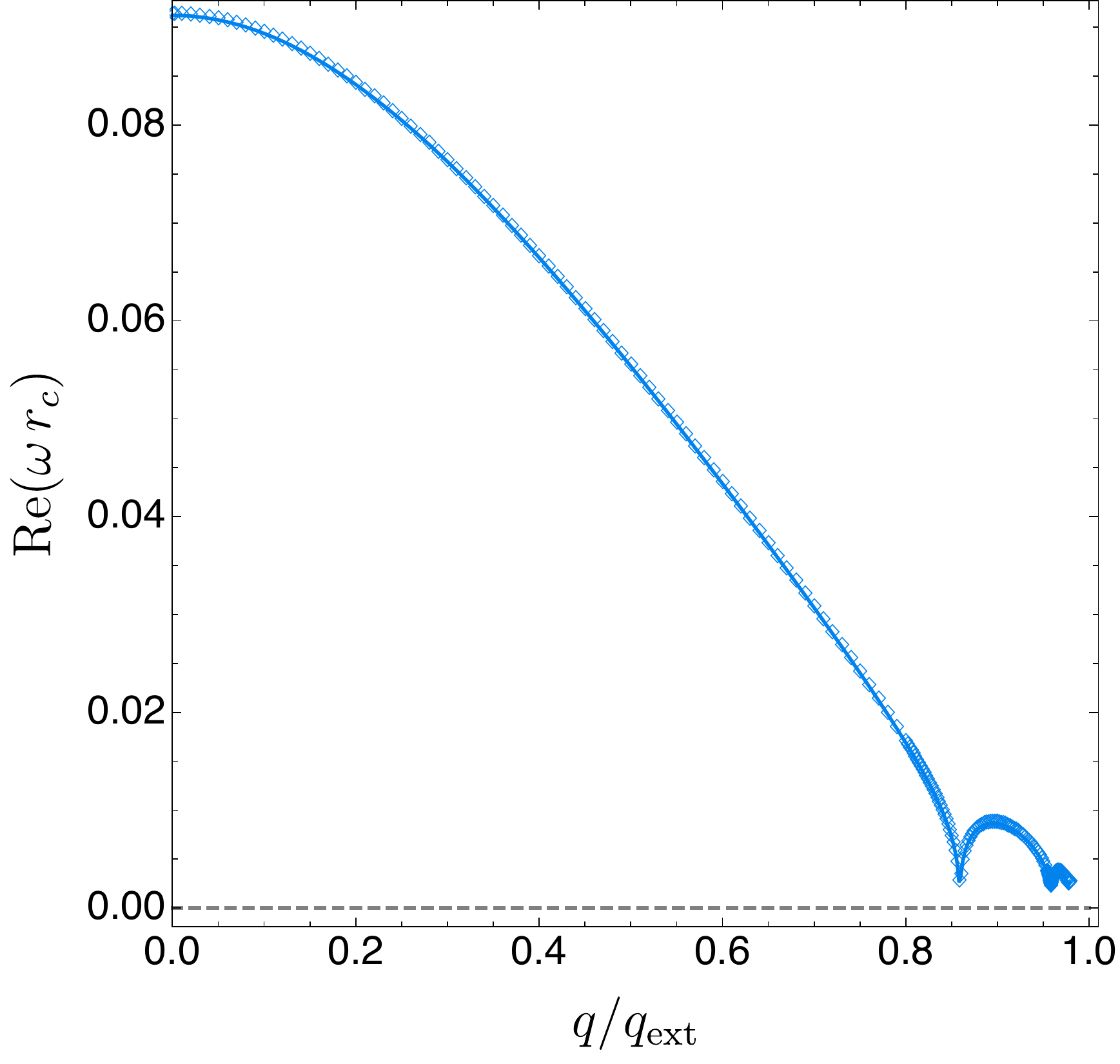} \hspace{0.3cm}
	\centering 	\includegraphics[width=0.49\textwidth]{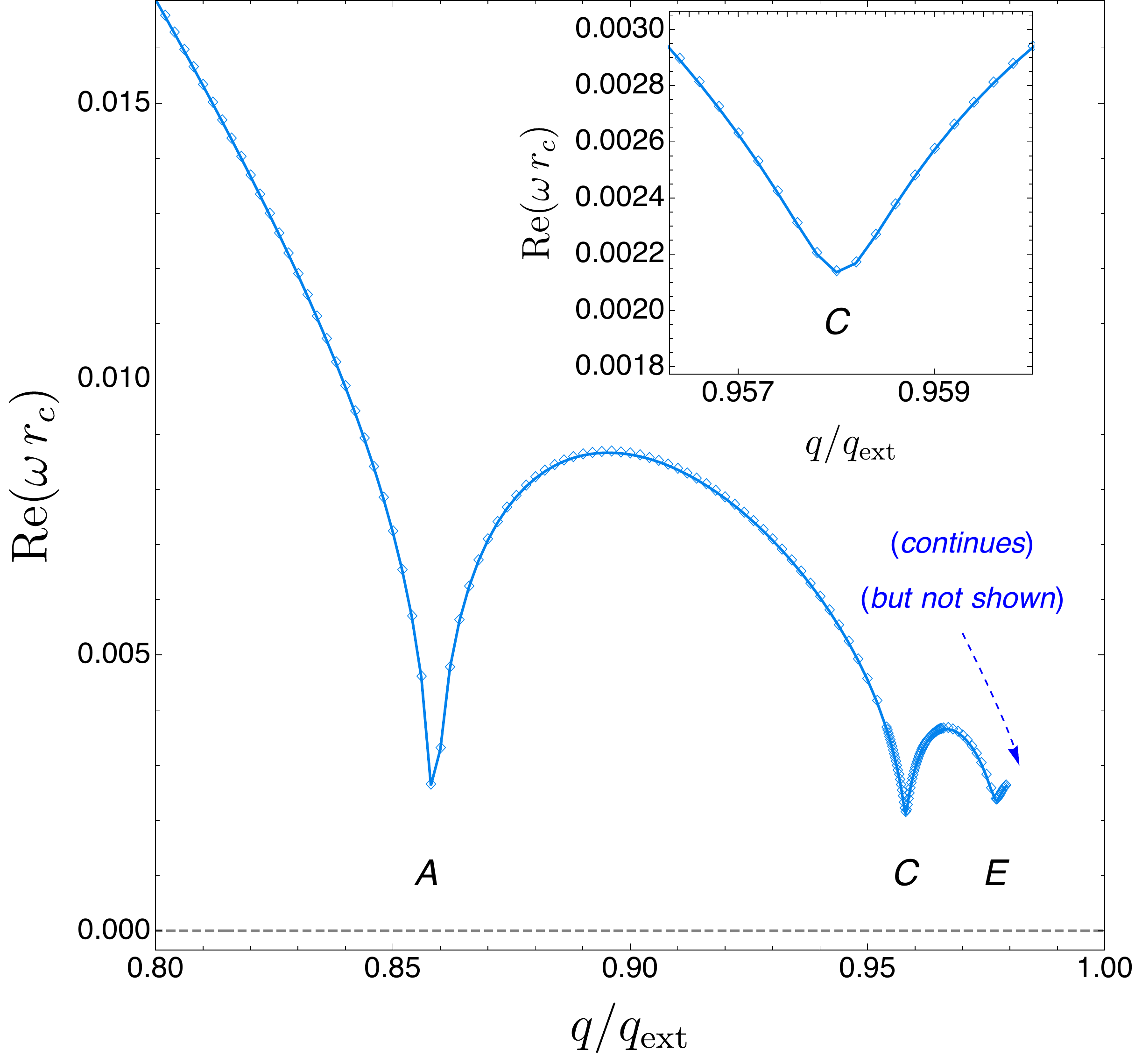} 
		\caption{Frequency spectrum  (real part) for $y_+=0.95$ and $n=5$, $\ell=2$ with zooms in the relevant regions where instability is present. The dark red and orange sections of Fig. \ref{fig:yp095Im} are {\it not} shown: they both have ${\rm Re}(\omega r_c)=0$.} 
		\label{fig:yp095Re}
 \end{figure} 
  This case distinguishes from the previous case mainly because the {\it bifurcations/mergers} (like $A$ and $C$ of Fig. \ref{fig:yp099Im}) cease to exist. Instead they are {\it replaced by crossovers} of quasinormal mode branches. Take the left panel of Figs. \ref{fig:yp095Im} and \ref{fig:yp095Re}. As before, the light blue lozenge ($\lozenge$) curve starts at $q=0$ and has  ${\rm Re}(\omega r_c)\neq 0 $. This curve extends all the way to extremality (although it becomes very difficult to compute the properties of this curve for $q/q_{\rm ext}$ above $0.98$) with an intricate structure best seen in the zoom plots of the right panel of Figs.  \ref{fig:yp095Im}  and  \ref{fig:yp095Re}. Indeed, in  the ${\rm Im}(\omega r_c)$ plot the light blue lozenge ($\lozenge$) curve has some zig-zagged regions that correspond to minima cusp points in the ${\rm Re}(\omega r_c)$ plot: these are  regions $A$, $C$, $E$, $\cdots$ in Figs.  \ref{fig:yp095Im}  and  \ref{fig:yp095Re}). These regions coincide with the points where the  dark red disk ($\bullet$) and  orange triangle ($\triangle$)  curves (that have both ${\rm Re}(\omega r_c)=0$) crossover the light blue lozenge ($\lozenge$)  curve (note that the red disk curve describing the most unstable mode is precisely the one already displayed in Fig. \ref{fig:severalyp} as a dark-red $\circ$). It is important to emphasize that $A$ and $C$ describe {\it crossover not  intersection} points (since the eigenfunctions of the branches that intersect are distinct and points $A$, $C$ and $E$ in the light blue $\lozenge$ curve have  ${\rm Re}(\omega r_c)\neq 0$ unlike in the $\bullet$ and $\triangle$ curves). Comparing with the situation of   Fig. \ref{fig:yp099Im} and  Fig. \ref{fig:yp099Re} we can say that the bifurcation/merger points $A$ and $C$ of Fig. \ref{fig:yp099Im} become crossover points in Fig. \ref{fig:yp095Im}. Further note that, although not shown in Fig. \ref{fig:yp095Im}, there should exist at least another family of modes to the right of the  orange $\triangle$ curve and ``parallel'' to it that should crossover the $\lozenge$ curve in region $E$. Finally, we should pointed out that the orange empty ($\triangle$) triangle modes of Fig. \ref{fig:yp095Im} are {\it not} the extension of the orange filled triangle ($\blacktriangle$) modes of Fig. \ref{fig:yp099Im}: this will become clear later when we discuss  the curves  $\blacktriangle$ and  $\triangle$ of  Fig. \ref{fig:q095Im}.
    
   \item Finally, we consider the case with $y_+=0.70$  which is a representative case of what happens with RNdS black holes that have intermediate and small values of $y_+$. The relevant frequency spectrum for this case is displayed in  Fig. \ref{fig:yp070Im} (imaginary part in left panel and real part in the right panel). This case distinguishes from the previous case because the structure of the spectrum is now very simple: the ziz-zag regions on the ${\rm Im}(\omega r_c)$ plots of Fig. \ref{fig:yp095Im}, {\it i.e.} the cusps in the ${\rm Re}(\omega r_c)$ plots of  Fig. \ref{fig:yp095Re}, have now flatten out completely and we simply have very simple crossovers $A$ and $C$ between the light blue lozenge ($\lozenge$) (which has  ${\rm Re}(\omega r_c)\neq 0$) and, respectively, the red disk ($\bullet$) curve that is unstable (already shown in Fig. \ref{fig:severalyp} as a dark-red $\bullet$) and the orange triangle ($\triangle$) curve.   
        \begin{figure}[ht]
	\centering 	\includegraphics[width=0.47\textwidth]{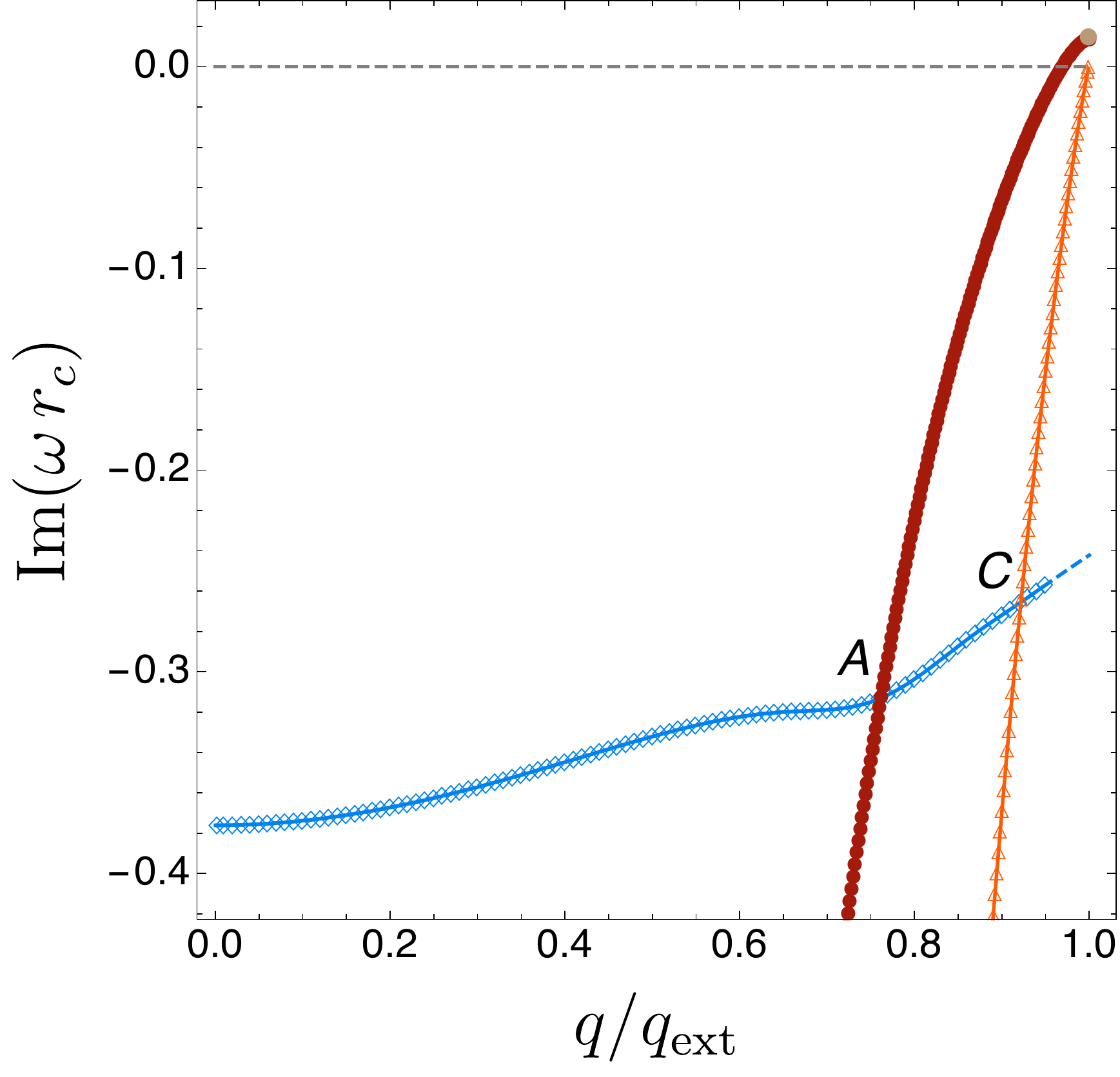} \hspace{0.3cm}
	\centering 	\includegraphics[width=0.46\textwidth]{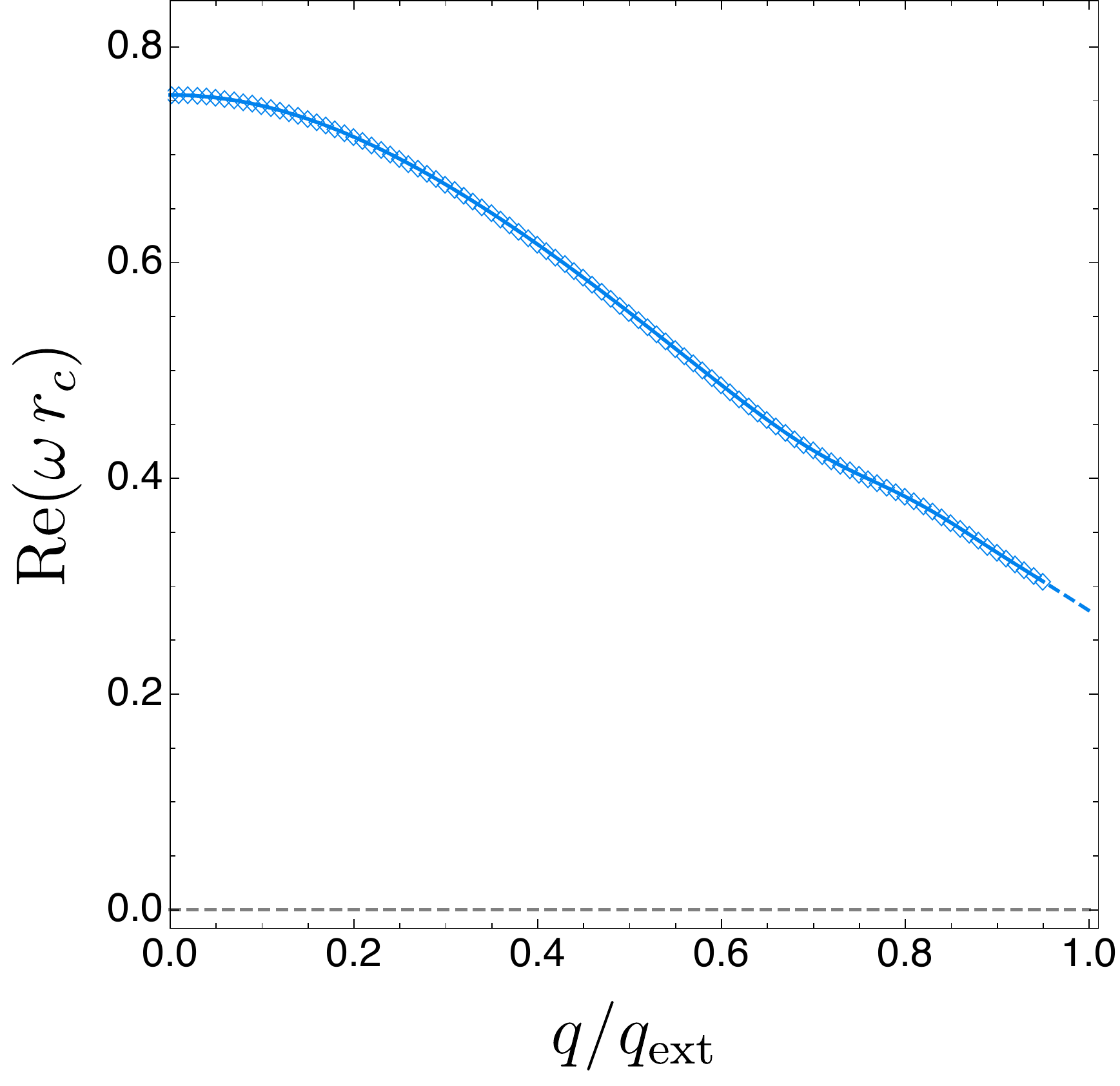} 
		\caption{Frequency spectrum (left panel: imaginary part, right panel: real part) for $y_+=0.70$ and $n=5$, $\ell=2$ with zooms in the relevant regions where instability is present. Right panel: Real part of the frequency. The dark red $\bullet$ and orange $\triangle$ branches  of the left panel have with ${\rm Re}(\omega r_c)=0$ and are {\it not} shown.} 
		\label{fig:yp070Im}
 \end{figure}

 \end{enumerate}
 
As observed before the RNdS family of black holes spans a 2-dimensional parameter space that we are taking to be the dimensionless ratios $y_+=r_+/r_c$ and $q/q_{\mathrm{ext}}$. 
In the sequence of Figs. \ref{fig:yp099Im}-\ref{fig:yp099Re}, \ref{fig:yp095Im}-\ref{fig:yp095Re} and \ref{fig:yp070Im} we have always fixed $y_+$ and analysed how the frequency spectrum changes as the charge of the RNdS black hole changes. 
Next, to have a 3-dimensional perspective of the system, we complement this analysis: we choose some relevant RNdS solutions with a fixed charge $q/q_{\mathrm ext}$ and discuss how their frequency changes as $y_+$ changes (we take again $n=5$ and $\ell=2$). Particularly illustrative cases that further help revealing the properties of the system are $q/q_{\mathrm{ext}}=0.95$, $q/q_{\mathrm{ext}}=0.88$ and $q/q_{\mathrm{ext}}=0.85$. The analysis of these three cases unveils the following properties:

 \begin{enumerate}[label=\arabic{*})]
 
    \begin{figure}[b]
	\centering 	\includegraphics[width=0.47\textwidth]{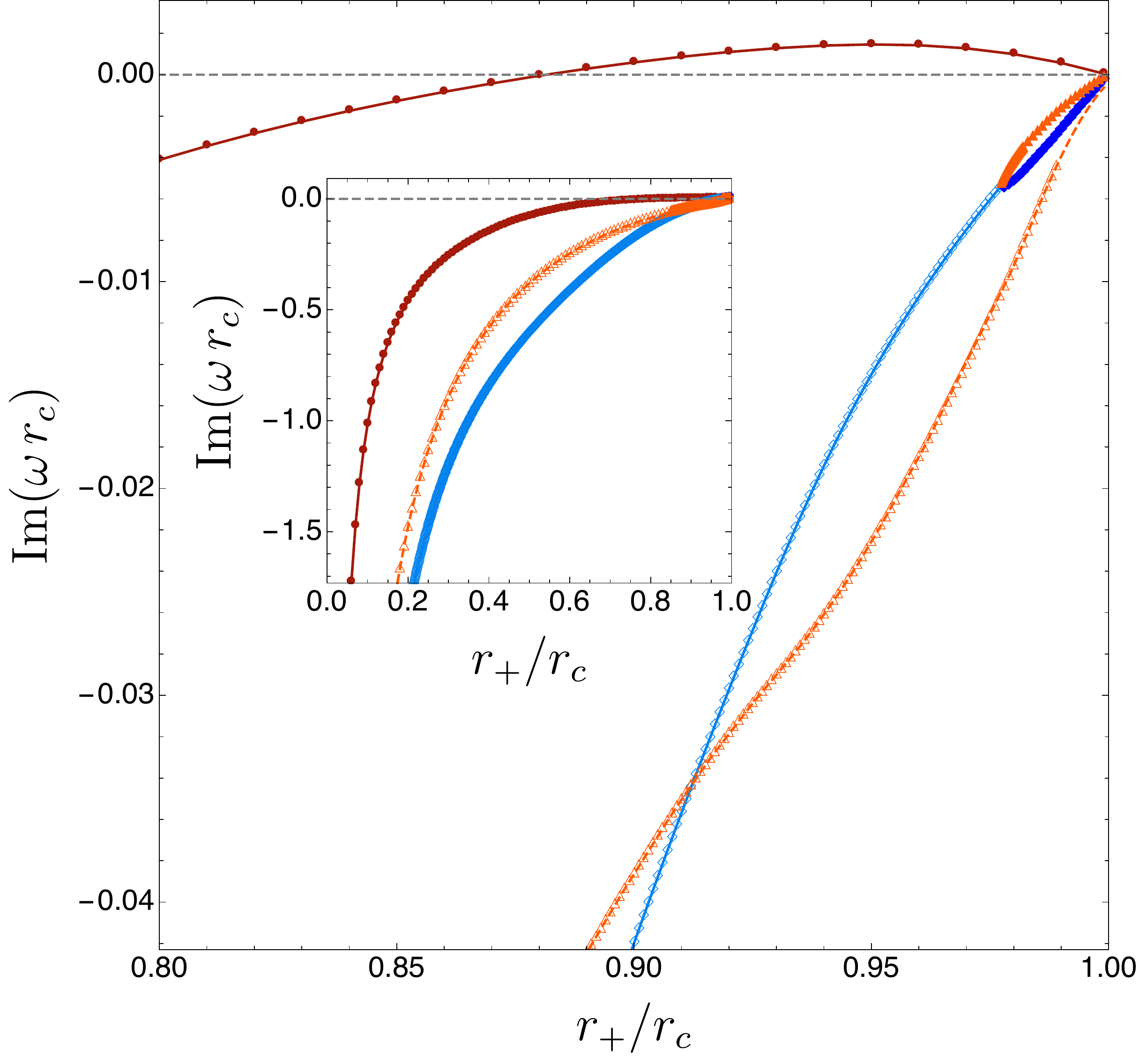} \hspace{0.3cm}
	\centering 	\includegraphics[width=0.485\textwidth]{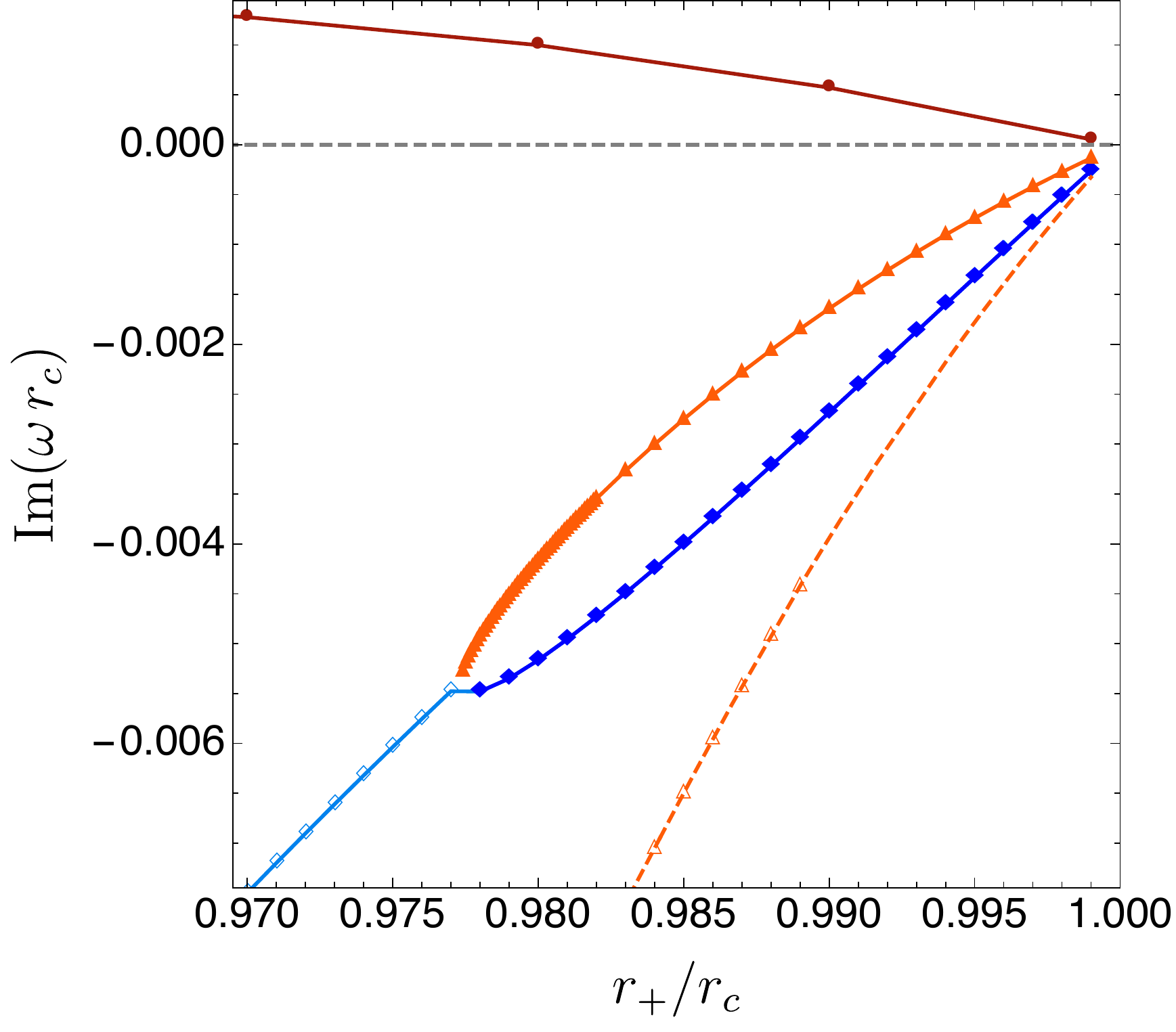} 
		\caption{Left and right panels: Frequency spectrum (imaginary part) for $q/q_{\mathrm ext}=0.95$ and $n=5$, $\ell=2$ with zooms in relevant regions. We use the same shape/colour code ($\bullet$, $\lozenge$,  $\blacktriangle$, $\blacklozenge$, $\triangle$) used in Figs \ref{fig:yp099Im}-\ref{fig:yp099Re}, \ref{fig:yp095Im}-\ref{fig:yp095Re} and \ref{fig:yp070Im}. } 
		\label{fig:q095Im}
 \end{figure} 
   \begin{figure}[ht]
	\centering 	\includegraphics[width=0.475\textwidth]{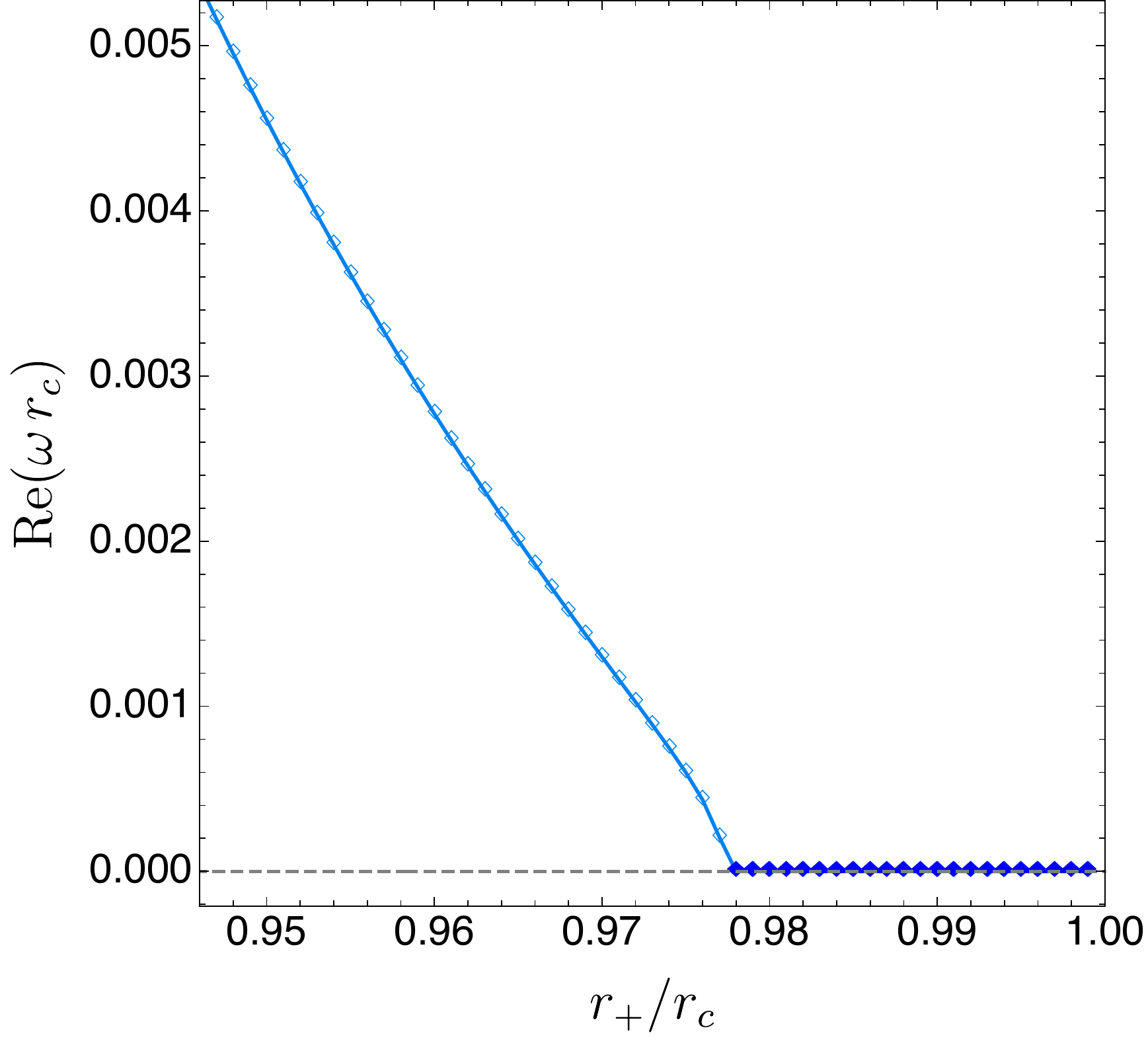} 
		\caption{Frequency spectrum (real part) for $q/q_{\mathrm ext}=0.95$ and $n=5$, $\ell=2$ with zooms in relevant regions. The dark red $\bullet$ and orange $\blacktriangle,\triangle$ branches of Fig. \ref{fig:q095Im} are {\it not} shown because they simply have ${\rm Re}(\omega r_c)=0$.} 
		\label{fig:q095Re}
 \end{figure} 
 
 \item We start with a RNdS black hole with $q/q_{\mathrm{ext}}=0.95$. In Fig. \ref{fig:q095Im} (imaginary part) and Fig. \ref{fig:q095Re} (real part) we show how the frequency $\omega r_c$ changes as we vary $y_+$. The curve on the top of Fig. \ref{fig:q095Im} with dark red disks ($\bullet$) is the branch that is unstable for large $y_+$. It confirms, as already seen in Fig. \ref{fig:onset}, that RNdS with $q/q_{\mathrm{ext}}= 0.95$ become unstable for $y_+ \gtrsim 0.881$ (see left panel of Fig. \ref{fig:q095Im}). Additionally, out of an infinite family of modes with  ${\rm Im}(\omega r_c)<0$ for any $y_+$, we further show the two families of modes that have the smallest  $|{\rm Im}(\omega r_c)|$. These are: i) the light blue $\lozenge$ branch (that has ${\rm Re}(\omega r_c)\neq 0$; see Fig. \ref{fig:q095Re}) that bifurcates at  $q/q_{\mathrm{ext}}\sim 0.977$ into two branches with ${\rm Re}(\omega r_c)=0$, namely the orange triangle ($\blacktriangle$)  branch  and the solid blue diamond ($\blacklozenge$)  branch, and ii) the orange empty triangle ($\triangle$) branch.  A zoom in of the region close to $y_+=1$ is displayed in the right panel of Fig. \ref{fig:q095Im}. This plot at constant  $q/q_{\mathrm{ext}}=0.95$ complements well those plots at constant $y_+$ shown previously (namely Figs. \ref{fig:yp099Im}, \ref{fig:yp095Im}, \ref{fig:yp070Im}). For example, in the right panel of Fig. \ref{fig:q095Im},  we identify the upper three points $\bullet$, $\blacktriangle$, $\blacklozenge$ with $y_+=0.99$. These three points are the three points with the same shape/colour code $\bullet$, $\blacktriangle$, $\blacklozenge$ with $q/q_{\mathrm ext}=0.95$ of Fig. \ref{fig:yp099Im} (which describes RNdS solutions with  $y_+=0.99$). As another example, in the left panel of Fig. \ref{fig:q095Im},  we identify the three points $\bullet$, $\blacklozenge$, $\triangle$ with $y_+=0.95$ that are the  three points  $\bullet$, $\blacklozenge$, $\triangle$ with $q/q_{\mathrm ext}=0.95$ of Fig. \ref{fig:yp095Im} (which describes RNdS solutions with  $y_+=0.95$). This is a good moment to pause and note, as observed in the end of the discussion of Fig. \ref{fig:yp095Im}, that the orange  $\blacktriangle$ (first introduced in Fig. \ref{fig:yp099Im}) and $\triangle$ modes  (first introduced in Fig. \ref{fig:yp095Im})  are indeed distinct. As a final example, in the left panel of Fig. \ref{fig:q095Im},  we further identify the three points $\bullet$, $\triangle$, $\blacklozenge$  with $y_+=0.70$ that are the  three points  $\bullet$,  $\triangle$, $\blacklozenge$ with $q/q_{\mathrm ext}=0.95$ of Fig. \ref{fig:yp070Im} (which describes RNdS solutions with  $y_+=0.70$).

  \item Next consider RNdS black holes with $q/q_{\mathrm{ext}}=0.88$. In Fig. \ref{fig:q088Im} (imaginary part) and Fig. \ref{fig:q088Re} (real part) we show the variation of the frequency as we vary $y_+$.  From Fig. \ref{fig:onset} (green $\blacksquare$ curve) such black holes are stable for all values of $y_+$. It follows that the curve on the top of Fig. \ref{fig:q088Im} with dark red disks ($\bullet$) has ${\rm Im}(\omega r_c)<0$ for any $y_+$: this is the family of modes that  become unstable but only for $q/q_{\mathrm{ext}}\gtrsim 0.881$. 
  Additionally, out of an infinite family of modes with  ${\rm Im}(\omega r_c)<0$ for any $y_+$, we further show the families of modes that have the smallest  $|{\rm Im}(\omega r_c)|$. This is the light blue $\lozenge$ branch (with ${\rm Re}(\omega r_c)\neq 0$; see Fig. \ref{fig:q088Re}) that bifurcates at  $q/q_{\mathrm{ext}}\sim 0.971$ into two branches with ${\rm Re}(\omega r_c)=0$, namely the blue diamond ($\blacklozenge$) branch  and the brown inverted triangle ($\blacktriangledown$) branch.  A zoom in of the region close to $y_+=1$ is displayed in the right panel of Fig. \ref{fig:q088Im}. This plot complements previous plots at constant $y_+$ (of Figs. \ref{fig:yp099Im}, \ref{fig:yp095Im}, \ref{fig:yp070Im}). For example, in the right panel of Fig. \ref{fig:q088Im},  we identify the upper three points $\bullet$, $\blacklozenge$, $\blacktriangledown$ with $y_+=0.99$ which are the three points $\bullet$, $\blacklozenge$, $\blacktriangledown$ with $q/q_{\mathrm ext}=0.88$ in the left panel of Fig. \ref{fig:yp099Im} (which describes RNdS solutions with  $y_+=0.99$; we do not show the green square family in  Fig. \ref{fig:q088Im}). As another example, still in the right panel of Fig. \ref{fig:q088Im},  we identify the two points $\bullet$, $\lozenge$ with $y_+=0.95$ that are the two points $\bullet$, $\lozenge$ with $q/q_{\mathrm ext}=0.88$ of Fig. \ref{fig:yp095Im} (which describes RNdS solutions with  $y_+=0.95$; we do not show the orange triangle family in Fig. \ref{fig:q088Im}). As a final example, in the left panel of Fig. \ref{fig:q088Im},  we also identify the  two points $\bullet$, $\lozenge$, this time with $y_+=0.70$, that are the two points $\bullet$, $\lozenge$ with $q/q_{\mathrm ext}=0.95$ of Fig. \ref{fig:yp070Im} (which describes RNdS solutions with  $y_+=0.70$).

    \begin{figure}[ht]
	\centering 	\includegraphics[width=0.47\textwidth]{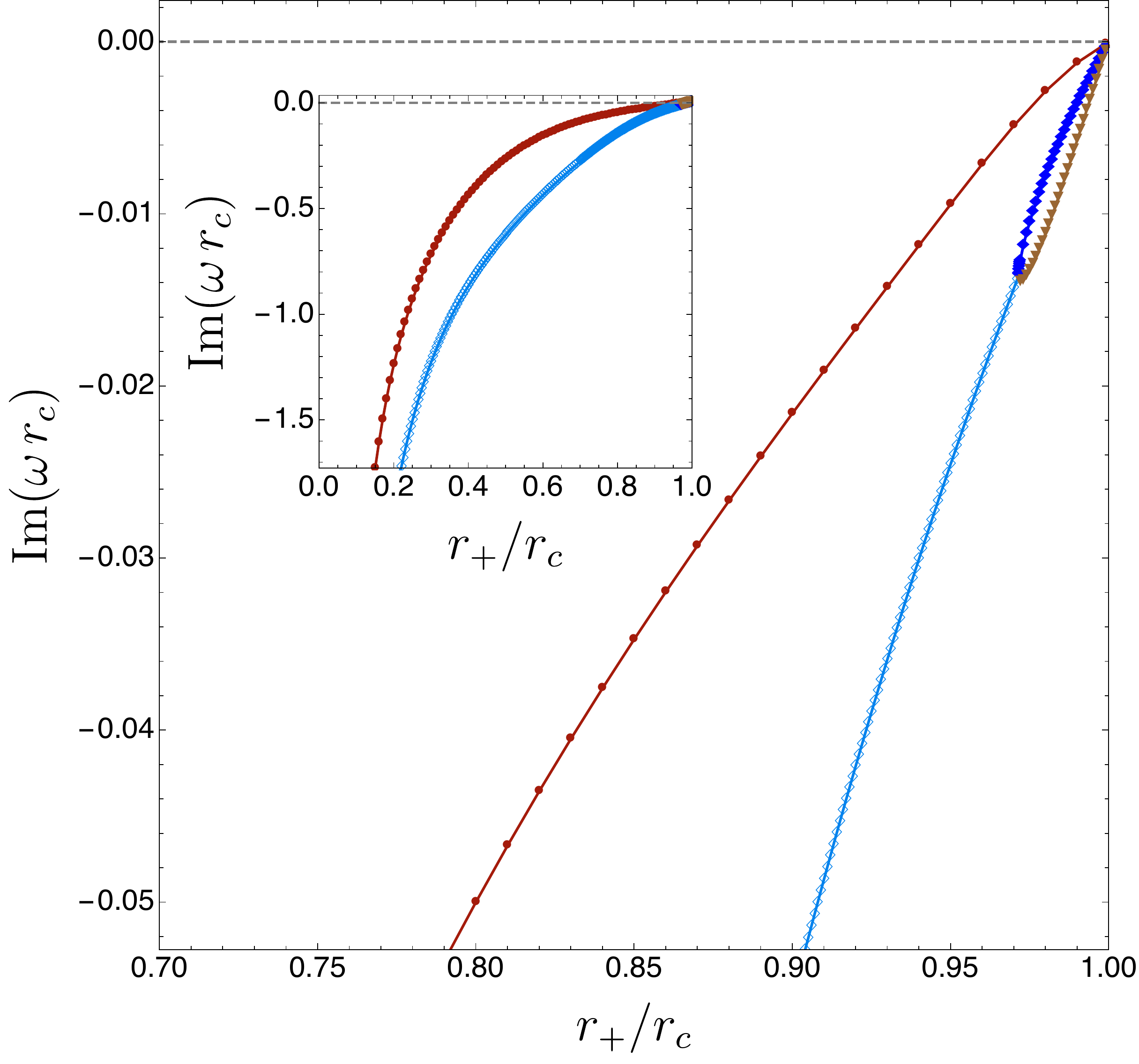} \hspace{0.3cm}
	\centering 	\includegraphics[width=0.485\textwidth]{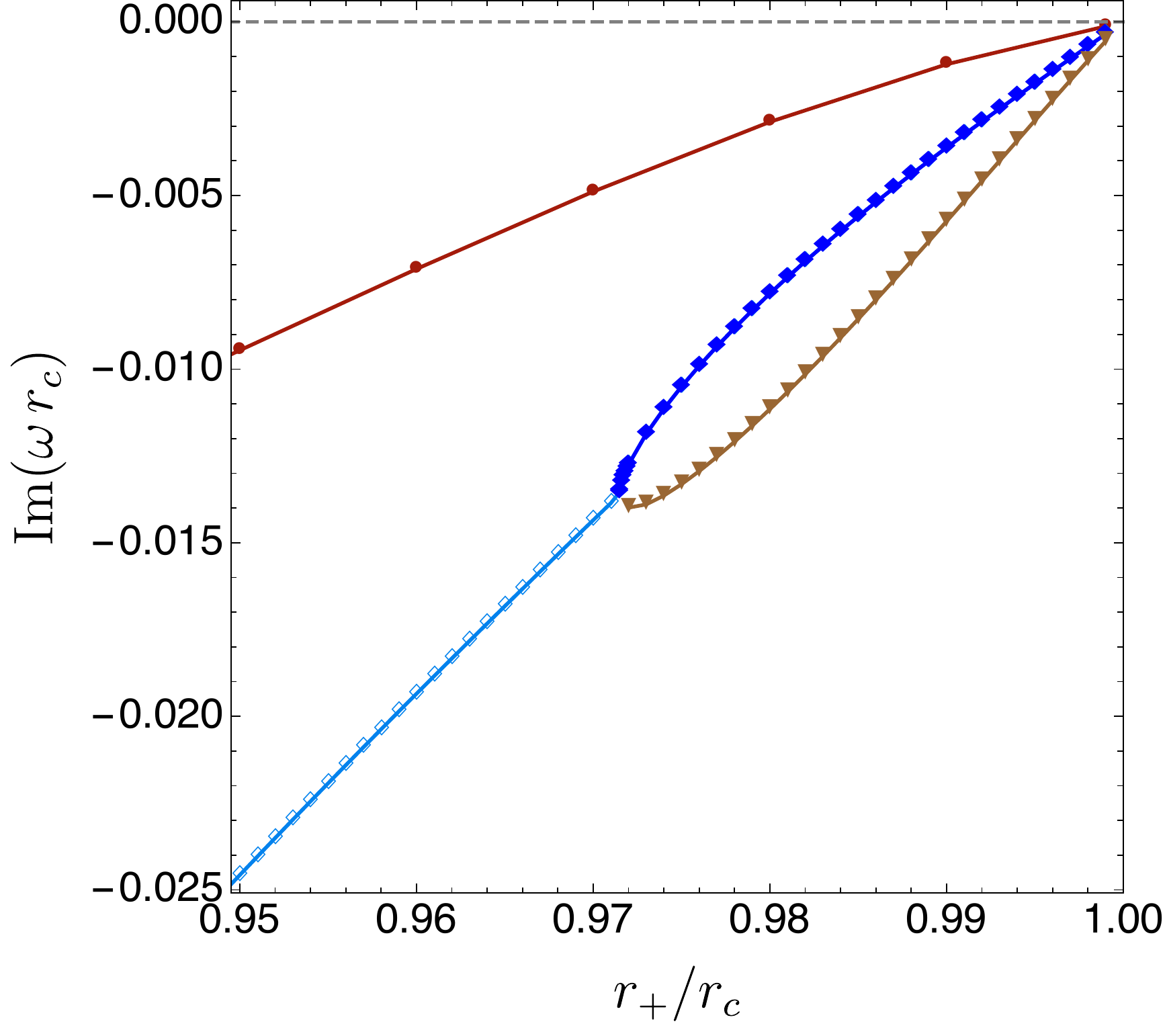} 
		\caption{Left and right panels: Frequency spectrum (imaginary part) for $q/q_{\mathrm ext}=0.88$ and $n=5$, $\ell=2$ with zooms in relevant regions. We use the same shape/colour code ($\bullet$, $\lozenge$,  $\blacklozenge$, $\blacktriangledown$) used in Figs \ref{fig:yp099Im}-\ref{fig:yp099Re}, \ref{fig:yp095Im}-\ref{fig:yp095Re} and \ref{fig:yp070Im}.} 
		\label{fig:q088Im}
 \end{figure} 
   \begin{figure}[ht]
	\centering 	\includegraphics[width=0.475\textwidth]{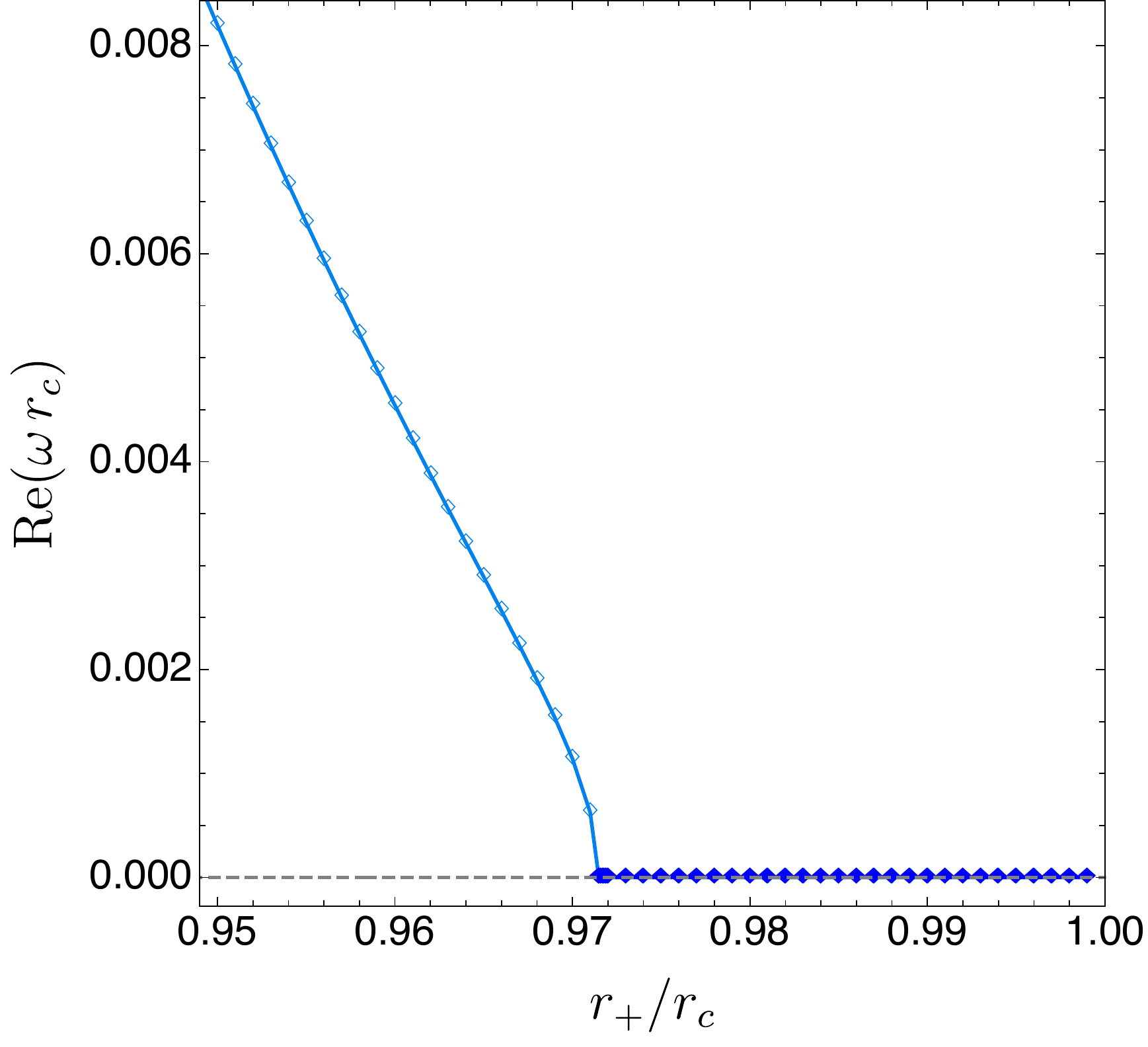} 
		\caption{Frequency spectrum (real part) for $q/q_{\mathrm ext}=0.88$ and $n=5$, $\ell=2$. The dark red $\bullet$ and brown inverted triangle $\blacktriangledown$  branches of Fig. \ref{fig:q088Im} are {\it not} shown because they simply have ${\rm Re}(\omega r_c)=0$.} 
		\label{fig:q088Re}
 \end{figure} 
 
  \item Finally, we considered RNdS black holes with $q/q_{\mathrm{ext}}=0.85$. In Figs.  \ref{fig:q085Im} and \ref{fig:q085Re} we display the three family of modes with the lowest $| {\rm Im}(\omega r_c)|$ for  this RNdS family. Such black holes are stable for all values of $y_+$ (see green $\blacksquare$ curve in Fig. \ref{fig:onset}).  In the main plot of the right panel,  we identify the blue $\lozenge$ point with $y_+=0.99$ which is also the blue $\lozenge$ point with $q/q_{\mathrm ext}=0.85$ in the left panel of Fig. \ref{fig:yp099Im} (which  describes solutions with  $y_+=0.99$). Also in the  main plot of the right panel,  we further identify the two points  (blue $\lozenge$ and dark red $\bullet$) with $y_+=0.95$ which are also the two points (blue $\lozenge$ and dark red $\bullet$) with $q/q_{\mathrm ext}=0.85$ in the right panel of Fig. \ref{fig:yp095Im} (which  describes solutions with  $y_+=0.95$). Finally, in the left panel, we  identify the two points (dark red $\bullet$ and blue $\lozenge$) with $y_+=0.7$ which are the two points (dark red $\bullet$ and blue $\lozenge$)  with $q/q_{\mathrm ext}=0.85$ of Fig. \ref{fig:yp070Im} (which describes solutions with  $y_+=0.70$). Note that one of the two branches (blue $\lozenge$ and dark red $\bullet$) has the lowest timescale depending on the window of $y_+$.
  
     \begin{figure}[ht]
	\centering 	\includegraphics[width=0.47\textwidth]{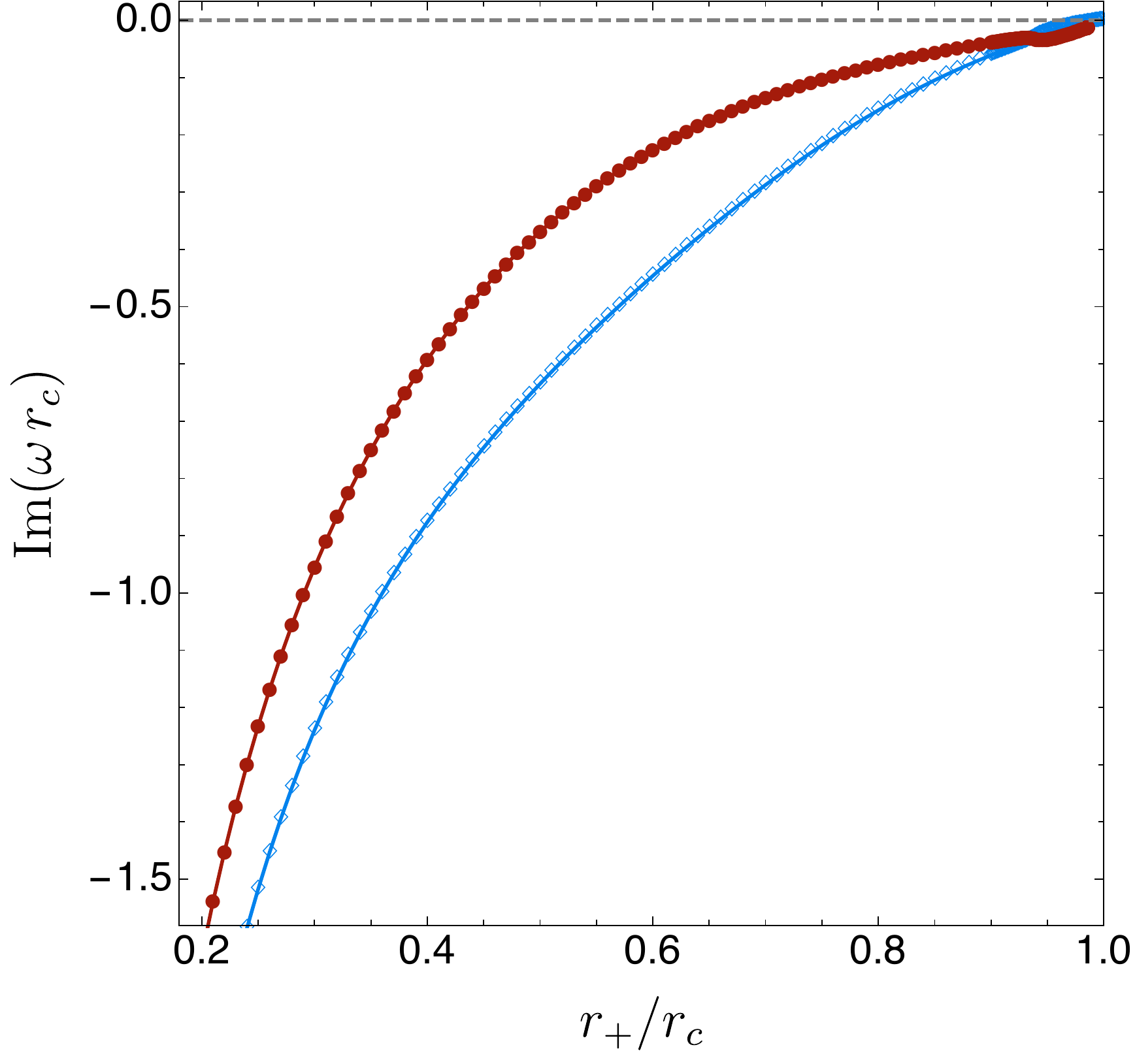} \hspace{0.3cm}
	\centering 	\includegraphics[width=0.485\textwidth]{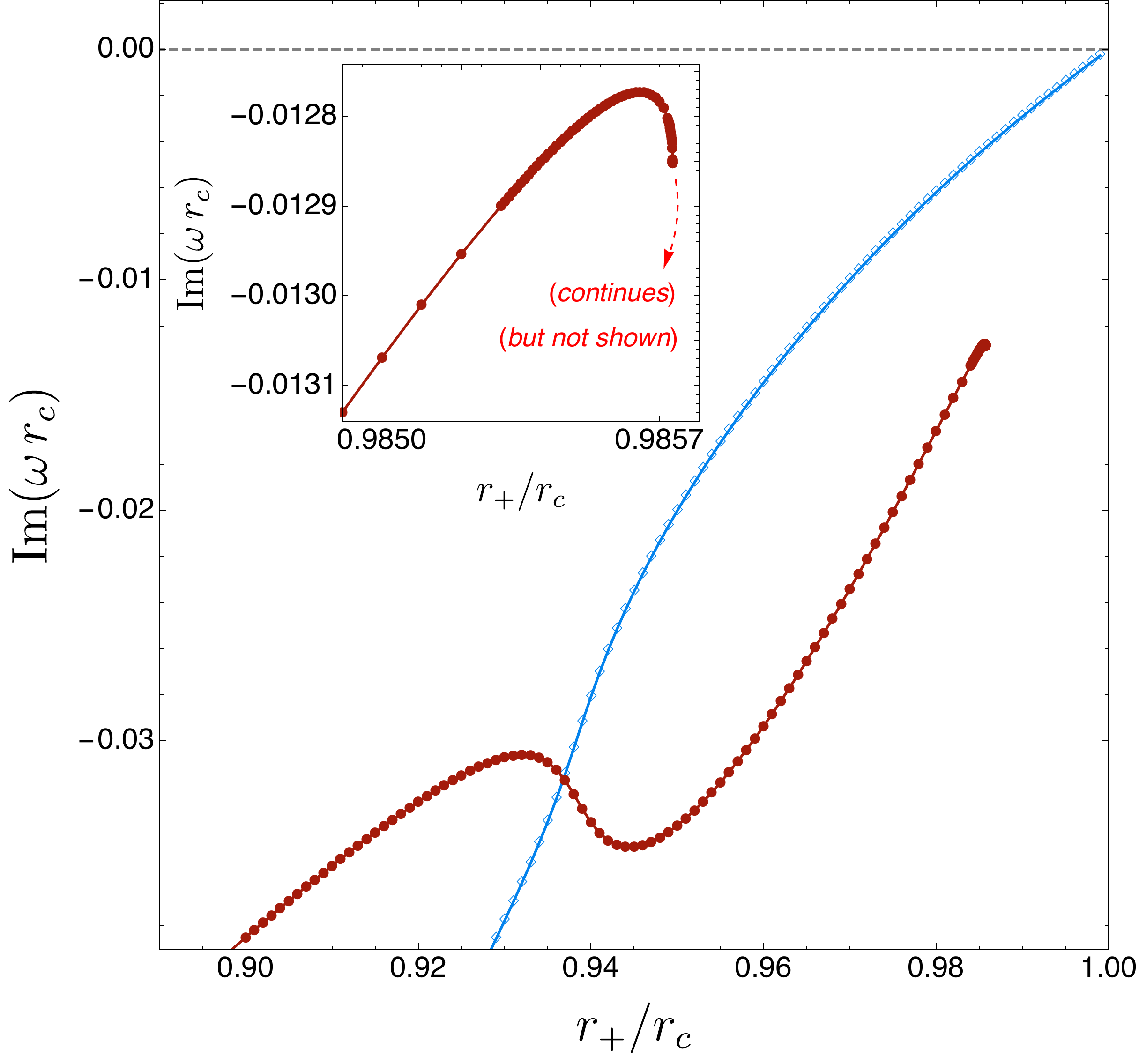} 
		\caption{Frequency spectrum (imaginary part) for $q/q_{\mathrm ext}=0.85$ and $n=5$, $\ell=2$ (left panel) with zooms in relevant regions (right panel).} 
		\label{fig:q085Im}
 \end{figure} 
   \begin{figure}[ht]
	\centering 	\includegraphics[width=0.475\textwidth]{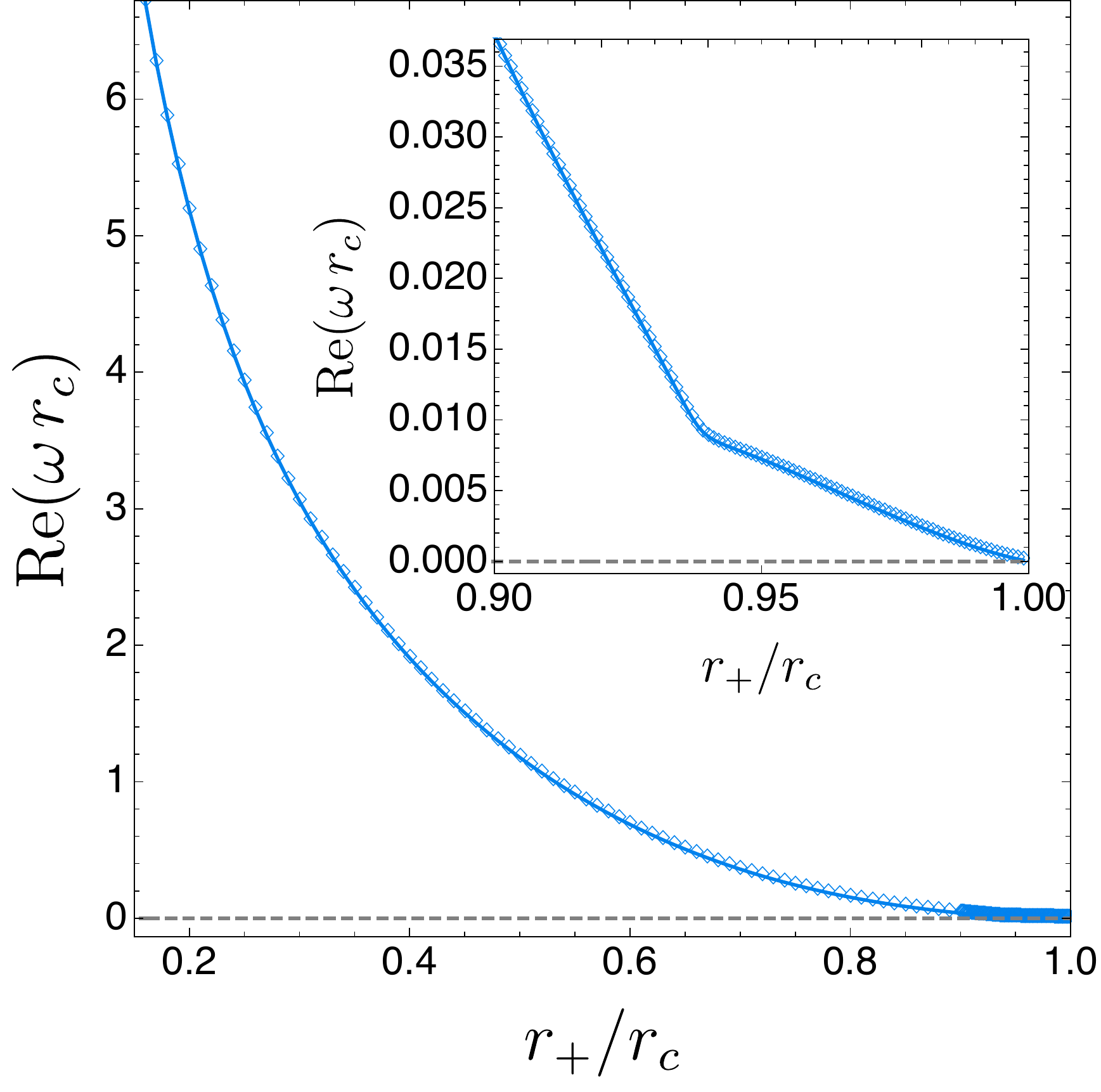} 
		\caption{Frequency spectrum (real part) for $q/q_{\mathrm ext}=0.85$ and $n=5$, $\ell=2$. The dark red $\circ$ and green  $\blacksquare$  branches of Fig. \ref{fig:q085Im} are {\it not} shown because they simply have ${\rm Re}(\omega r_c)=0$.} 
		\label{fig:q085Re}
 \end{figure} 
 
 \end{enumerate}
 
\section{Discussion} 
\label{sec:discussion} 

We believe that our study reveals key aspects of the gravitational instability of Reissner-Nordstr\"om de Sitter black holes originally found in \cite{Konoplya:2008au} and further studied in \cite{Cardoso:2010rz,Tanabe:2015isb}. The fundamental novel results that we establish  can be summarized as follows:
\begin{itemize}
\item RNdS black holes are unstable when $d\geq 6$ (see {\it e.g.} Fig. \ref{fig:extreme}). For $d\geq 7$ this instability was first established in \cite{Konoplya:2008au,Cardoso:2010rz,Tanabe:2015isb}. In addition, we find it is also present for $d=6$ (although with a much lower timescale).

\item We have established the physical origin of the instability: it is present because in the near-horizon limit of extremal RNdS,  the unstable gravitational modes effectively behave as a massive scalar field whose mass violates the AdS$_2$ Breitenl\"ohner-Freedman bound (if and only if $d\geq 6$; see Fig. \ref{fig:NH}). By continuity, the instability then extends away from extremality.

\item The instability criterion of the previous item is known as the Durkee-Reall conjecture \cite{Durkee:2010ea} later proved by Hollands and Wald \cite{Hollands:2014lra} when $\Lambda\leq 0$. Our findings provide good numerical evidence for the Durkee-Reall conjecture with de Sitter asymptotics ($\Lambda>0$), as already argued in \cite{Hollands:2014lra}. It would be interesting to formally prove this result for $\Lambda>0$ using the methods of \cite{Hollands:2014lra}.

\item Our results also confirm that the Durkee-Reall instability criterion \cite{Durkee:2010ea,Hollands:2014lra} provides a sufficient but not necessary condition for instability. For example, for extremal black holes the instability criterion typically predicts instability only for certain windows of the dimensionless horizon radii ratio $r_+/r_c$ (see Fig. \ref{fig:NH}) but we find that actually the instability is present for any value of this ratio (at extremality and for $d\geq 6$): see  {\it e.g.} Fig. \ref{fig:extreme}. In particular, this also means that there is no critical minimum value of $y_+$ for the existence of the instability unlike it was claimed in \cite{Konoplya:2013sba}.

\item RNdS black holes are parametrized by the dimensionless ratios $y_+\equiv r_+/r_c$  and $q/q_{\mathrm{ext}}$. For a given dimension $n$ and $y_+$, we have found  the onset charge $q^{\mathrm{onset}}/q_{\mathrm{ext}}$ above which RNdS becomes unstable: see Fig. \ref{fig:onset}.

\item In addition to finding the instability timescale at extremality (Fig. \ref{fig:extreme})  and the onset charge  (Fig. \ref{fig:onset}) we have spanned the 2-dimensional parameter space of RNdS to produce the 3-dimensional plot of  Fig. \ref{fig:3D} that displays the instability timescale as a function of $y_+\equiv r_+/r_c$  and $q/q_{\mathrm{ext}}$. In particular, as best seen in Fig. \ref{fig:severalyp}, after its onset the instability first increases as we approach extremality until it reaches a maximum near but before extremality. Then its strength decreases slightly as we approach further and reach extremality.
  
\item The instability is present for $d\geq 6$ for the harmonic mode $\ell=2$. Typically, the instability then tends to get weaker and even disappear as the harmonic number $\ell$ grows (see Fig. \ref{fig:harmonic}). For example: in $d=6,7$ only the $\ell=2$ mode is unstable; in $d=8,9$ $\ell=3$ is also unstable; and in $d=10,11$  $\ell=2,3,4$ modes are unstable but not $\ell\geq 5$.  

\item If we follow the unstable modes to regions of the parameter space where the instability shuts down and if we include the second and/or third families of quasinormal modes with lowest $|{\rm Im}(\omega r_c)|$ we find an intriguing network of mode bifurcations/mergers that seems to be absent in black holes with $\Lambda\leq 0$. This spectrum seems so unique and intriguing that we dedicated a special study to it. In the sequence of Figs. \ref{fig:yp099Im}-\ref{fig:yp099Re}, \ref{fig:yp095Im}-\ref{fig:yp095Re} and \ref{fig:yp070Im} we have fixed representative values of $y_+$ and analysed how the frequency spectrum changes as the charge of the RNdS black hole varies. On the other hand, in Figs. \ref{fig:q095Im}-\ref{fig:q095Re}, \ref{fig:q088Im}-\ref{fig:q088Re} and \ref{fig:q085Im}-\ref{fig:q085Re} we fixed some values of $q/q_{\mathrm{ext}}$ and changed $y_+$. 
  
\end{itemize}

There are quite a few natural extensions of our work. In particular, it would be interesting to frame the quasinormal mode structure we found in this manuscript in light of the three families of quasinormal modes used to study the strong cosmic censorship conjecture \cite{penrose} for initial data close to RNdS black holes \cite{Cardoso:2017soq,Costa:2017tjc,Dias:2018ynt,Dias:2018etb,Dafermos:2018tha,Hod:2018dpx,Dias:2018ufh,Dias:2019ery,Zhang:2019nye,Gim:2019rkl,Liu:2019lon}. We should note, however, that once the instability sets in, a novel black hole geometry is likely to form \cite{DiasSantos:2020}. Strong cosmic censorship should then be studied by investigating how slowly generic perturbations decay around the new hypothetical background geometry. Away from extremality, we should be able to settle this issue by studying the quasinormal mode spectrum of the relevant RNdS black hole. 

\subsection*{Acknowledgments}

We would like to thank V.~Cardoso, R.~A.~Konoplya and A.~Zhidenko for reading an earlier version of this manuscript.
OJCD is supported by the STFC Ernest Rutherford Grant No. ST/K005391/1, and by the STFC ``Particle Physics Grants Panel (PPGP) 2016" Grant No. ST/P000711/1. J. E. S. is supported in part by STFC grants PHY-1504541 and ST/P000681/1. J. E. S. also acknowledges support from a J. Robert Oppenheimer Visiting Professorship. This work used the DIRAC Shared Memory Processing system at the University of Cambridge, operated by the COSMOS Project at the Department of Applied Mathematics and Theoretical Physics on behalf of the STFC DiRAC HPC Facility (www.dirac.ac.uk). This equipment was funded by BIS National E- infrastructure capital grant ST/J005673/1, STFC capital grant ST/H008586/1, and STFC DiRAC Operations grant ST/K00333X/1. DiRAC is part of the National e-Infrastructure.


\bibliographystyle{JHEP}
\bibliography{ref_ds}{}

\end{document}